\documentclass{aa}
\usepackage[varg]{txfonts}
\usepackage{graphicx,multirow,aalongtable,natbib,epstopdf}
\bibpunct{(}{)}{;}{a}{}{,}
\graphicspath{{./figure/}}

\begin{document}

\title{The relationship between radio power at 22 and 43 GHz \\
and black hole properties of AGN in elliptical galaxies}

\author{Songyoun Park\inst{1,2}
\and Bong Won Sohn\inst{2}\thanks{\email{bwsohn@kasi.re.kr}}
\and Sukyoung K. Yi\inst{1}}

\institute{Department of Astronomy and Yonsei University Observatory, Yonsei University, Seoul 120-749, Republic of Korea
\and Korea Astronomy and Space Science Institute, Daejeon 305-348, Republic of Korea}

\date{Received 18 February 2013 / Accepted 31 July 2013}

\abstract {We investigate the relationship between radio power and properties related to active galactic nuclei (AGNs). Radio power at 1.4 or 5 GHz, which has been used in many studies, can be affected by synchrotron self-absorption and free-free absorption in a dense region. On the other hand, these absorption effects get smaller at higher frequencies. Thus, we performed simultaneous observations at 22 and 43 GHz using the Korean VLBI Network (KVN) radio telescope based on a sample of 305 AGN candidates residing in elliptical galaxies from the overlap between the Sloan Digital Sky Survey (SDSS) Data Release 7 and Faint Images of the Radio Sky at Twenty-Centimeters (FIRST). About 37\% and 22\% of the galaxies are detected at 22 and 43 GHz, respectively. Assuming no flux variability between the FIRST and KVN observation, spectral indices were derived from FIRST and KVN data and we found that over 70\% of the detected galaxies have flat or inverted spectra, implying the presence of optically thick compact regions near the centres of the galaxies. Core radio power does not show a clear dependence on black hole mass at either low (1.4 GHz) or high (22 and 43 GHz) frequencies. However, we found that the luminosity of the [OIII] $\lambda$5007 emission line and the Eddington ratio correlate with radio power more closely at high frequencies than at low frequencies. This suggests that radio observation at high frequencies can be an appropriate tool for unveiling the innermost region. In addition, the luminosity of the [OIII] $\lambda$5007 emission line and the Eddington ratio can be used as a tracer of AGN activity. Our study suggests a causal connection between high frequency radio power and optical properties of AGNs.
}

\keywords{galaxies: active -- galaxies: nuclei -- galaxies: elliptical and lenticular, cD -- radio continuum: galaxies}
\titlerunning{Radio Power of AGN in Elliptical Galaxies}
\authorrunning{Park, Sohn, \& Yi}

\maketitle

\section{Introduction}
Active galactic nuclei (AGNs) are thought to be important in galaxy formation and evolution. It is widely accepted that supermassive black holes (SMBHs) are ubiquitous in the centres of galaxies \citep{1998Natur.395A..14R}. Moreover, several studies have found that black hole mass correlates well with bulge mass \citep{1995ARA&A..33..581K, 1998AJ....115.2285M, 2002MNRAS.331..795M, 2003ApJ...589L..21M} and velocity dispersion \citep{2000ApJ...539L...9F, 2000ApJ...539L..13G, 2001ApJ...547..140M}. These findings suggest that central black holes are closely linked to their host galaxies. According to the black hole paradigm \citep{1977MNRAS.179..433B}, nuclear activity is powered by mass accretion onto the black holes at the centres of galaxies, and its radiation is observed over a wide range of wavelengths. In particular, radio emission has been characterized based on intriguing phenomena related to nuclear compact cores and extended features such as jets and lobes. However, in spite of great efforts to understand the features and physics of AGN, we still lack a detailed explanation of how the energy related to an accretion disk is produced, collimated, and propagated or of how supermassive black holes grow.

A number of earlier studies have attempted to determine the relationship between radio power and AGN properties such as black hole mass, emission line luminosity, and the Eddington ratio. Some authors have reported a correlation between radio luminosity and black hole mass \citep{1998MNRAS.297..817F, 2000ApJ...543L.111L, 2001ApJ...551L..17L, 2002MNRAS.336L..38J, 2003MNRAS.340.1095D, 2004MNRAS.353L..45M}. However, other studies have not found this relationship \citep{2002ApJ...564..120H, 2002ApJ...579..530W, 2002ApJ...576...81O, 2003MNRAS.342..889S}. Furthermore, some studies have reported that a relationship between radio power and optical narrow-line luminosity \citep{1989ApJ...336..702B, 1991Natur.349..138R, 1998MNRAS.298.1035T}, while other studies have questioned this relationship \citep{2001ApJ...555..650H, 2005MNRAS.362...25B}. Regarding black hole accretion rates, \cite{2007AJ....133.2187D} reported that radio power correlates with the Eddington ratio, which was later confirmed by \cite{2012MNRAS.421.1569B}. However, the radio luminosity used in previous studies is mostly observed at low frequencies, i.e. 1.4 or 5 GHz.

The continuum radio emission associated with AGN originates from synchrotron radiation emitted by charged particles moving relativistically in magnetic fields. Compact core components show low frequency cutoff in power-law radio spectra. Two physical processes (synchrotron self-absorption and free-free absorption) can explain the origin of this convex spectra \citep{1973ranp.book.....P, 1979rpa..book.....R}. The compact radio sources are optically thick because of synchrotron self-absorption \citep{2008evn..confE...5O}. Free-free absorption occurs in dense ionized regions such as in obscuring tori \citep{2000PASJ...52..209K, 2002astro.ph..4054K}.

The aim of this study is to investigate the relationship between radio power and black hole properties of AGNs. Because of the absorption mechanisms mentioned above, low frequencies such as 1.4 or 5 GHz can be inefficient for measuring the intrinsic radio power of AGNs. By using high frequencies which are optically thin to these absorptions, we are able to measure radio power from core components. Futhermore, simultaneous observations at 22 and 43 GHz will eliminate variability uncertainties in spectral index.

Several observations at either 22 or 43 GHz have been conducted on elliptical or radio galaxies. We used optical emission line measurements to represent the black hole properties. We compared optical emission lines of host galaxies with radio power. We focused on three different measures of AGN activity: velocity dispersion ($\sigma$), luminosity of the [OIII]$\lambda$5007 emission line, and black hole accretion rate. In order to exclude Doppler boosted beaming sources, we removed all broad line region galaxies from our sample.

This paper is organized as follows. In Sect. 2, we present a description of sample selection. In Sect. 3, we introduce the Korean VLBI Network (KVN) radio telescope system, and describe the observations and data reduction. In Sect. 4, we present the results of radio observations and the Sloan Digital Sky Survey (SDSS) data analysis. In Sect. 5, we discuss the connection between radio power and AGN activity.

Throughout this paper, we assume the cosmological parameters with $\Omega_{m} = 0.3$, $\Omega_{\Lambda} = 0.7$, and H$_{0} = 70$ km s$^{-1}$ Mpc$^{-1}$.

\section{Sample}
We constructed a sample for the KVN single-dish observation. We made use of the database of the two major surveys in radio and optical wavelengths: FIRST and SDSS Date Release 7 (DR7). We selected galaxies for which photometric, spectroscopic, and radio data are available, as described below, in order to compare radio power at frequencies of 1.4 GHz (FIRST) and 22 GHz (KVN) with the optical properties (SDSS) of AGNs in elliptical galaxies.

\subsection{The Sloan Digital Sky Survey}
The Sloan Digital Sky Survey \citep{2000AJ....120.1579Y, 2002AJ....123..485S} includes imaging and spectroscopic data obtained using a dedicated 2.5-meter telescope at Apache Point Observatory. The telescope has a distortion-free field of view of 3$^{\circ}$. The imaging is carried out using a drift scan camera, which has a pixel size of 0.396$''$ and an exposure time of 53.9 seconds at each filter. The central wavelengths of each band are $u$, $g$, $r$, $i$, and $z$ = 3551 \AA, 4686 \AA, 6165 \AA, 7481 \AA, and 8931 \AA. The magnitude limits for 95\% completeness are 22.0, 22.2, 22.2, 21.3, and 20.5, respectively. The coverage of the spectrograph is from 3800 \AA$\,$ to 9200 \AA$\,$. The SDSS fibres have a finite diameter of 3$''$, which corresponds to projected separation of about 0.6 kpc at z $=$ 0.01 and 3.5 kpc at z $=$ 0.06. Therefore, each fibre covers the central region of the galaxies. We used the seventh data release, which is the final public data of the SDSS-II project that was completed in July 2008 \citep{2009ApJS..182..543A}. 

\subsection{The FIRST Survey}
The FIRST project \citep{1995ApJ...450..559B} was originally proposed to obtain radio counterparts to the Palomar Observatory Sky Survey (POSS I), which covers optical sources in the northern sky. However, the plan was modified to cover the northern Galactic gap, which corresponds to the SDSS. The observations have been carried out up to the present using the Very Large Array (VLA). The array consists of 27 radio antennae in a Y-shaped arrangement, and each antenna has a diameter of 25 metres. The FIRST survey has been carried out using synthesis snapshot mode in the B configuration. The angular resolution is 5$''$, which corresponds to projected separation of about 1.0 kpc at z $=$ 0.01 and 5.8 kpc at z $=$ 0.06. The 1.4 GHz receiver system (L band) has 2 (IFs) $\times$ 2 (polarizations) $\times$ 7 (frequency) channels. The width of each channel is 3 MHz, and the total bandwidth is 168 MHz. The central frequencies are 1365 MHz and 1435 MHz. The system temperature in the L band is 37 to 75 K. The catalogue employed in the present work was released in 2010 and contains about 800,000 sources. The sky coverage of the FIRST survey closely matches that of the SDSS. The FIRST survey has a detection threshold of 0.75 mJy with a typical rms of 0.15 mJy \citep{1997ApJ...475..479W}.

\subsection{Sample selection}
We set up a sample with the following criteria. A redshift range from 0.01 to 0.06 and an absolute magnitude cut of M$_{r}$ $<$ -19.4 were employed for volume limitation. With the optically selected galaxies, we inspected visually if there were radio sources within a certain angular distance. Finally, we classified Seyferts, LINERs, and star-forming galaxies via a BPT diagram \citep{1981PASP...93....5B}.

\subsubsection{Photometric data}
We extracted 131,038 galaxies from SDSS DR7 with a redshift criteria of 0.01 $<$ z $<$ 0.06. The lower limit of redshift was set to 0.01 in order to avoid saturation. The upper limit was set to 0.06 because it was difficult to classify the morphology of galaxies at SDSS image quality beyond a redshift of 0.06. We estimated the magnitude of the galaxies, taking into account Galactic foreground extinction and \textit{k}-correction. Galactic extinction was corrected using the dust maps provided by \cite{1998ApJ...500..525S}. \cite{2003AJ....125.2348B} and \cite{2007AJ....133..734B} provided a method to conduct \textit{k}-correction for each galaxy via SED fitting.

The apparent magnitude limit for the main galaxy sample of SDSS is 17.77 in Petrosian r band \citep{2002AJ....124.1810S}. This limit corresponds to the absolute magnitude M$_{r}$ $=$ -19.4 at z $=$ 0.06. Thus, we chose galaxies with M$_{r}$ $<$ -19.4 in order to build a volume-limited sample.

\subsubsection{Spectroscopic data}
The SDSS project released the spectroscopic data for the main galaxy sample. \cite{2011ApJS..195...13O} provided measurements of stellar kinematics and improved line measurements for the SDSS galaxy spectra using pPXF \citep[penalized pixel-fitting;][]{2004PASP..116..138C} and GANDALF \citep[Gas AND Absorption Line Fitting;][]{2006MNRAS.366.1151S} code. They claimed to achieve this improvement by applying more realistic spectral templates and internal gas extinction \citep[see][]{2011ApJS..195...13O}.

Based on the \cite{2011ApJS..195...13O} database, we checked the strength of four emission lines: [NII] $\lambda$6583, [OIII] $\lambda$5007, H$\alpha$ $\lambda$6563, and H$\beta$ $\lambda$4861, all of which were used for spectroscopic diagnostics \citep{1981PASP...93....5B}. In this study, we selected galaxies with amplitude-over-noise (A/N) $>$ 2 in all four emission lines. This relatively low A/N criterion was used to build an unbiased elliptical galaxy sample. We only included galaxies with very strong emission lines if the A/N cut is too high, since elliptical galaxies usually have weak or no emission lines. This makes it possible to miss genuine AGN host galaxies with low-ionized emission lines.

\subsubsection{Radio data}
We matched the optically selected galaxies with FIRST data. The SDSS-FIRST galaxies are dominated by core sources \citep{2002AJ....124.2364I}. Over 90\% of radio sources have optical counterparts within a radius of 3$''$ \citep{2002AJ....124.2364I, 2008AJ....136..684K}. However, elliptical galaxies are frequently associated with extended radio structures such as jets or lobes. Therefore, to find radio counterparts and to derive flux from FIRST data, we applied an adjustable matching radius based on the structure of the radio source. We initially applied a searching radius of 10$''$ for all optically selected galaxies. If a galaxy had a single or multiple radio detections within 10$''$, then the matching radius was lengthened to 30$''$. Additional searches were then made between 10$''$ and 30$''$. In the same manner, when extra radio detections were found within 30$''$, the matching radius was increased to 60$''$.

These SDSS-FIRST selected galaxies were visually inspected using both SDSS optical images and FIRST radio images. Visual inspection allowed us to identify true radio emissions associated with the host galaxies. Moreover, we classified the morphology of those galaxies and selected elliptical galaxies. We reviewed whether there are more detections with the radius $>$ 60$''$ and whether the radio sources are related to the selected galaxies with radio jets, bridges, or extended lobes. When there were multiple detections within this searching radius, we only integrated the fluxes related to the elliptical galaxies without contamination. We manually measured core flux and extended flux separately from FIRST images. In the end, 4587 galaxies have been identified with radio detections, and about 8\% (374/4587) of them are classified as elliptical galaxies. The radio sources of the elliptical galaxies show either only core (81\%) or extended (19\%) structures. We found three sources without a core, but the high frequency flux shows flat spectra that will be compact, young, variable candidates.

\subsubsection{Emission line diagnostics}
To categorize galaxies as star-formation dominated or AGN dominated galaxies, we utilized BPT emission line diagnostics \citep{1981PASP...93....5B}. The distinct location of AGNs and star-forming galaxies on the BPT diagram is determined by the difference in the main excitation mechanisms. Emission lines in star-forming galaxies are formed by massive and hot stars, while AGN emission lines are powered by a SMBH accretion disk. The emission lines of O and B stars in star-forming galaxies have an upper limit on the intensities of the collisionally excited lines relative to recombination lines such as H$\alpha$ and H$\beta$. On the other hand, AGN emission lines are ionized by harder continua with a greater fraction of high energy photons, such as the power-law spectrum \citep{1981PASP...93....5B, 1997iagn.book.....P}. Several pairs of narrow emission line ratios are used for the BPT diagram. We used [NII]/H$\alpha$ and [OIII]/H$\beta$, which are strong and easy to detect (Kewley et al. 2001; hereafter Ke01). 

\begin{figure} 
\resizebox{\hsize}{!}{\includegraphics{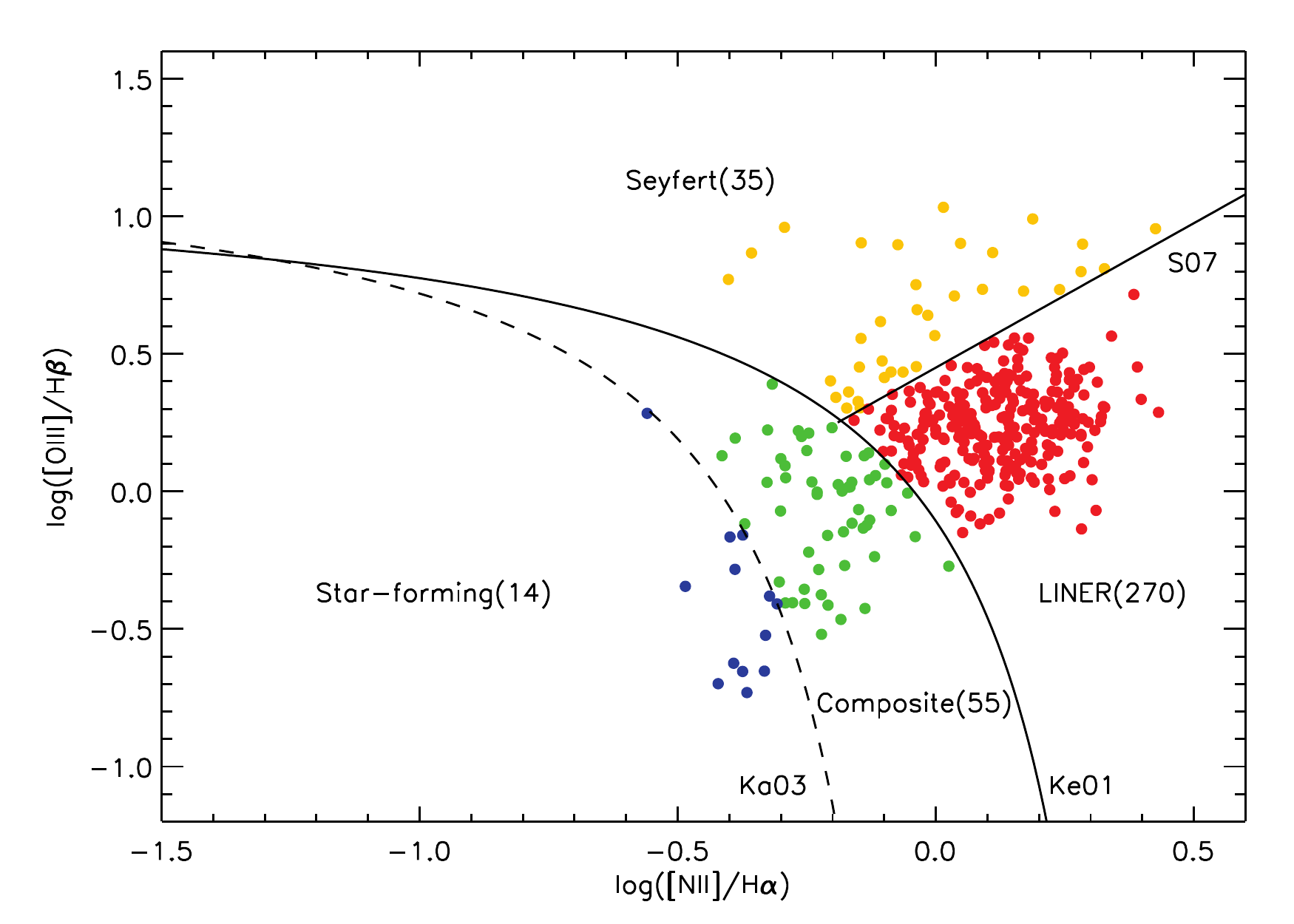}}
\caption{Emission line diagnostic diagram \citep{1981PASP...93....5B}. Only galaxies with A/N $>$ 2 are presented. Of our galaxy sample, about 88\% are classified as AGN dominated galaxies, and most of them are LINERs.}
\label{BPT}
\end{figure}

Figure~\ref{BPT} shows a BPT diagram of our sample. The galaxies are divided into four classes by three demarcation lines: star-forming, composite, Seyfert, and LINER \citep[low-ionization nuclear emission-line regions;][]{1980A&A....87..152H}. The dashed curve is derived empirically using the SDSS galaxies 
(Kauffmann et al. 2003; hereafter Ka03). The solid curve is determined by using both photoionization and stellar population synthesis models (Ke01). 
\cite{2007MNRAS.382.1415S} (hereafter S07) defined the empirical demarcation line dividing Seyfert and LINER. Because their morphological criteria are consistent with our sample, we adopted this categorization:
\begin{eqnarray}
 log\,\left(\frac{[OIII]}{H \beta}\right) = \frac{c}{(log\,\left([NII]/H\alpha\right))^{a}-b} + d .
 \label{eqBPT}
\end{eqnarray}

If galaxies are located below the dashed curve (Ka03), the emission lines are dominated by star formation. Galaxies above the solid curve (Ke01) represent AGN dominated galaxies. This solid curve is the theoretical upper limit of star formation activity. The S07 division of LINER and Seyfert is based on their [OIII] luminosities \citep{2006MNRAS.372..961K}. The region labelled composites lies between the lines dividing the star-formation dominated and AGN dominated galaxies in the diagram. The composite objects undergo both star formation and AGN activity at the central region of galaxies (Ke01). These three lines can be formulated by Eq.~(\ref{eqBPT}). The coefficients used to construct this BPT diagram are listed in Table~\ref{tab:BPTcons}. We finally constructed a sample of 305 AGN-hosting elliptical galaxies for the KVN 22 and 43 GHz observation.

\begin{table}[b]
 \begin{center}
  \caption[Coefficients of demarcation lines of Figure~\ref{BPT}]
  {Coefficients of demarcation lines}
  \begin{tabular}{l c c c c}
  \hline \hline
 \multicolumn{1}{c}{} & a & b & c & d \\
 \hline
 Ka03 & 1 & 0.50 & 0.61 & 1.30 \\
 Ke01 & 1 & 0.47 & 0.61 & 1.19 \\
 S07 & -1 & 0 & 1.05 & 0.45 \\
   \hline
\end{tabular}
\label{tab:BPTcons}
\end{center}
\end{table}

\begin{table*}
  \caption{Summary of sample selection process}
  \centering
  \begin{tabular}{l r r}
  \hline \hline
 \multicolumn{1}{l}{Criterion} & Explanation & No. of Galaxies\\
 \hline
 Redshift &$0.01 < z < 0.06$ & 131,038 \\
 Absolute magnitude & M$r < -19.4$ & 83,800 \\
 Emission line strength & A/N $>$ 2 & 52,964 \\
 Matching radii & 10$''$, 30$''$ and 60$''$ & 4587 \\
 Visual inspection & Elliptical galaxies & 374 \\
 BPT diagram & AGN (Seyfert \& LINER) & 305 \\
   \hline
\end{tabular}
\label{tab:sample}
\end{table*}

\begin{figure*}
\resizebox{\hsize}{!}{\includegraphics{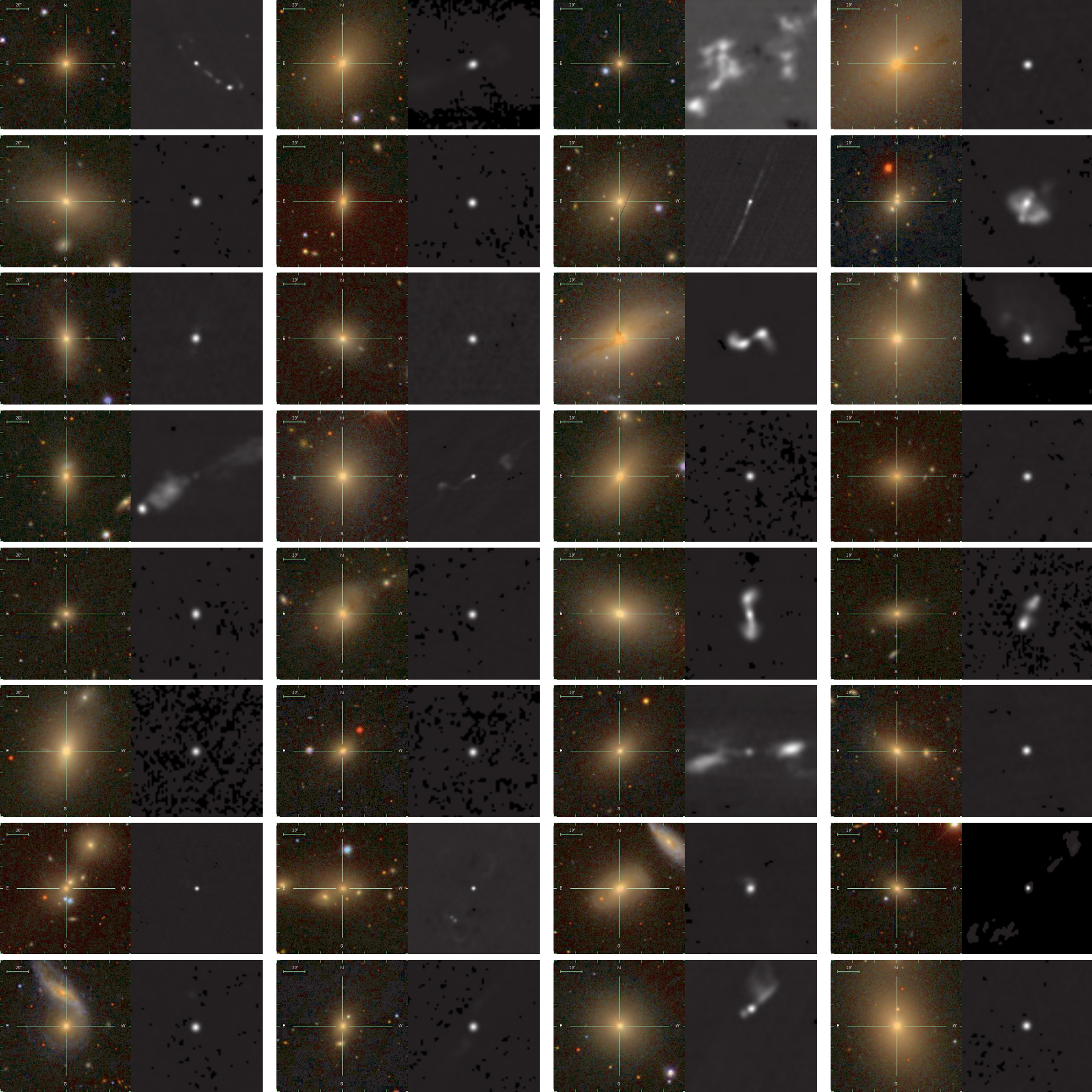}}
\caption{Optical and radio images of detected galaxies at 22 or 43 GHz. Each image is 2 $'\times$2 $'$.}
\label{ima_det}
\end{figure*}

\section{Observations and data reduction}

\subsection{Observations}
The Korean VLBI Network is the first dedicated millimetre-wave VLBI facility in East Asia \citep{2004evn..conf..281K}. The KVN is comprised of three 21-m Cassegrain shaped radio telescopes located in Seoul (Yonsei University), Ulsan (University of Ulsan), and Jeju (Jeju International University). The distinctive feature of the KVN radio telescope is simultaneous multi-frequency observation. 

We performed single-dish observations at 22 (K band) and 43 GHz (Q band) simultaneously using the KVN radio telescopes at Yonsei and Ulsan observatories. We carried out observations over 400 hours from November 2010 through March 2011. The bandwidth of the continuum backend was 512 MHz and typical system temperatures $T_{A}^*$ were 86 and 121 K at 22 and 43 GHz, respectively. All observations were carried out using cross-scan (CS) mode. One CS unit consists of two one-dimensional on-the-fly (OTF) in azimuth and elevation. This is an ideal compromise to achieve both sufficient accuracy and a high duty cycle. The pointing offset of the telescope, which influences the flux density measurements, can be determined and corrected simultaneously during observations \citep{2004PhDT}. Tables~\ref{tab:KVN_obs} and ~\ref{tab:KVN_obs_log} present specifications and a summary of the KVN single-dish observations.

We checked the pointing accuracy using closeby strong radio sources \citep{2011PASP..123.1398L} while holding the pointing offset at less than 4$''$. We observed a 3C 286, primary calibrator, in order to convert the measured antenna temperature into flux density. For each target, we repeated cross-scanning for 100 units. All the sources were observed when they are located between 30$^{\circ}$ and 65$^{\circ}$ in elevation. We selected 100 out of the 305 galaxies in our sample.

\begin{table*}
 \centering
  \caption[Specifications of KVN single-dish observations]{Specifications of KVN single-dish observations}
  \begin{tabular}{l c c}
  \hline \hline
    \multicolumn{3}{c}{Specification} \\
  \hline
 Frequency & 21.7 GHz (K) & 42.4 GHz (Q) \\
 Typical System Temperature ($T_{A}^*$) & 86 K (K) & 121 K (Q) \\
 Integration Time (on source) & 400 s (K) & 200 s (Q)\\
 Half-Power Beam Width (HPBW) & 130$''$ (K) & 65$''$ (Q) \\
 Observational Mode & \multicolumn{2}{c}{Az-El Cross Scan Mode}  \\ 
 Bandwidth & \multicolumn{2}{c}{512 MHz} \\
 Primary Calibrator & \multicolumn{2}{c}{3C 286} \\
 \hline
 \end{tabular}
\label{tab:KVN_obs}
\end{table*}

\begin{table*}
 \centering
  \caption[KVN single-dish observation log]{KVN single-dish observation log}
  \begin{tabular}{c c c c c}
  \hline \hline
    \multicolumn{1}{c}{Session} & Date & Site & Time & No. of Galaxies \\
  \hline
 1 & 2010 Nov. 25 - 2010 Nov. 30 & Ulsan & 6h-20h (LST) & 18 \\
 2 & 2010 Dec. 25 - 2011 Jan. 03 & Ulsan & 6h-20h (LST) & 33 \\
 3 & 2011 Feb. 14 - 2011 Feb. 26 & Yonsei & 9h-18h (LST) & 30 \\
 4 & 2011 Mar. 21 - 2011 Mar. 30 & Ulsan & 6h-20h (LST) & 19 \\
 \hline
 \end{tabular}
\label{tab:KVN_obs_log}
\end{table*}

\subsection{Data reduction}
We reduced the observational data using CLASS software from the GILDAS package (http://iram.fr/IRAMFR/GILDAS/). Cross-scan data were divided into two groups with respect to line types (azimuth and elevation) in order to perform pointing correction in each direction. Linear baseline fitting was applied to estimate the noise level. We then performed Gaussian fitting to measure antenna temperature and pointing offset. 

In CS mode, pointing offset in one direction affects the antenna temperature measurement of the other direction. We corrected this with 
\begin{eqnarray}
{(T_{A}^*)_{corrected,i}} = (T_{A}^*)_{observed,i} \cdot exp \left[ 4\,ln2 \cdot \left( \frac{x_{j}^{2}}{\theta_{j}^{2}} \right) \right] , \, i \neq j \, ,
\label{eq:pointcor}
\end{eqnarray}
where $(T_{A}^*)_{observed}$, $x$, and $\theta$ represent the observed antenna temperature in \textit{i} direction, pointing deviation, and the HPBW of Gaussian fit in \textit{j} direction, respectively, where \textit{i} and \textit{j} are azimuth or elevation \citep{2004PhDT}. The typical pointing accuracy of the KVN antenna is 5$''$ \citep{2011PASP..123.1398L}. If the pointing offset was 5$''$, the antenna temperature was corrected by 1\% at 22 GHz and 3\% at 43 GHz with Eq.~(\ref{eq:pointcor}).

After pointing correction, we calculated the source flux density and rms noise level. In order to determine the flux density, we used 3C 286. The fluxes of 3C 286 were 2.64 Jy at 22 GHz and 1.5 Jy at 43 GHz, as obtained from Mars observations (Sohn et al. in preparation). A typical conversion factor for KVN is around 12 Jy K$^{-1}$ at 22 GHz and 13 Jy K$^{-1}$ at 43 GHz. We observed 86 out of the 100 galaxies. The remaining 14 sources were excluded owing to the bad weather conditions.

\section{Results}

\subsection{KVN detections}
Table~\ref{tab:KVN_detection} presents the results of the KVN single-dish observations. Of 86 AGN in elliptical galaxies, 32 (37\%) are detected at 22 GHz ($>$ 3$\sigma$ level). Furthermore, among the 32 detected galaxies, 19 (22\%) are detected at 43 GHz. However, no galaxies are detected only at 43 GHz. The lower detection rate at 43 GHz can be explained in one or more of three ways. First, system temperature is higher at 43 GHz. Second, the integration time at 43 GHz is only half of the time at 22 GHz. As the integration time increases, the noise level decreases, which makes it possible to detect fainter sources. Third, sources are usually fainter at 43 GHz. 

Assuming that the flux densities at both frequencies are the same, in other words the spectrum is flat, and that the rms noise level and integration time at 43 GHz are applied, 15 out of the 32 galaxies detected at 22 GHz would be detected. The results of observation are given in Table~\ref{tab:KVN_detection}. Figure~\ref{ima_det} shows the optical and radio images of 32 detected galaxies. For non-detected galaxies, we set 3$\sigma$ as the upper limit of flux density, where $\sigma$ is the observation rms noise level.

\begin{figure}
\resizebox{\hsize}{!}{\includegraphics{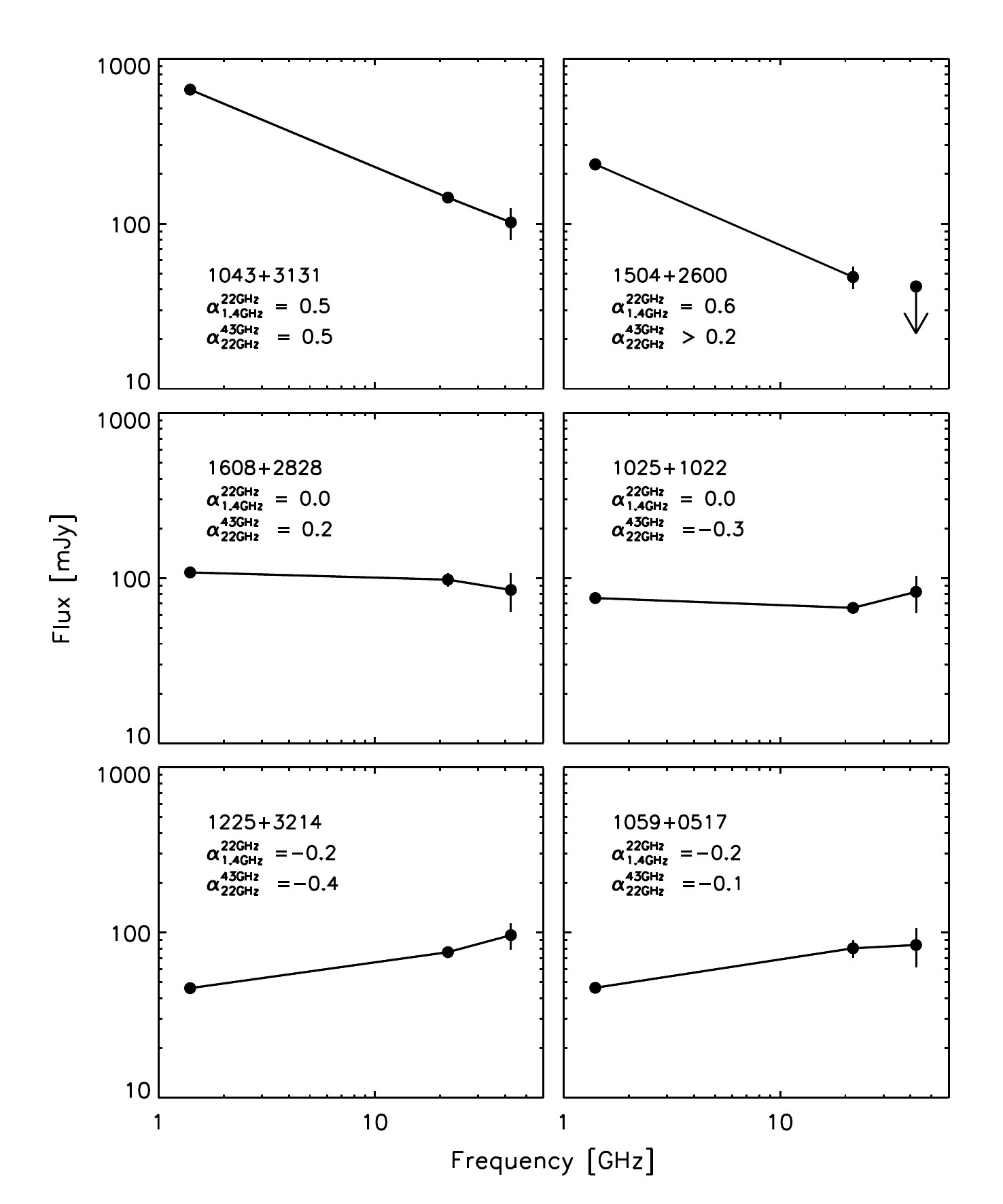}}
\caption[Examples of spectral index plots]
{Examples of spectral index plots. Spectral slopes of steep spectra are shown in the top panels, flat spectra in the middle panels, and inverted spectra in the bottom panels. An arrow pointing downwards indicates the upper limit of flux density.}
\label{spindex}
\end{figure}

\subsection{Spectral index distribution}
The spectral index $\alpha$ is defined as $S_{\nu} \propto \nu^{-\alpha}$, where $S_{\nu}$ is the flux density of a source and $\nu$ is the observing frequency. We estimated spectral indices, as given in Table~\ref{tab:KVN_detection}, using the results from the FIRST survey at 1.4 GHz and simultaneous KVN single-dish observations at 22 and 43 GHz. For non-detected galaxies, the spectral index was determined from the upper limit of flux density. Figure~\ref{spindex} presents examples of radio spectra at frequencies ranging from 1.4 GHz to 43 GHz. We classified the galaxies into three groups according to spectral index: steep ($\alpha > 0.5$), flat ($0 < \alpha < 0.5$), or inverted ($\alpha < 0$) spectra. 

\begin{figure}
\resizebox{\hsize}{!}{\includegraphics{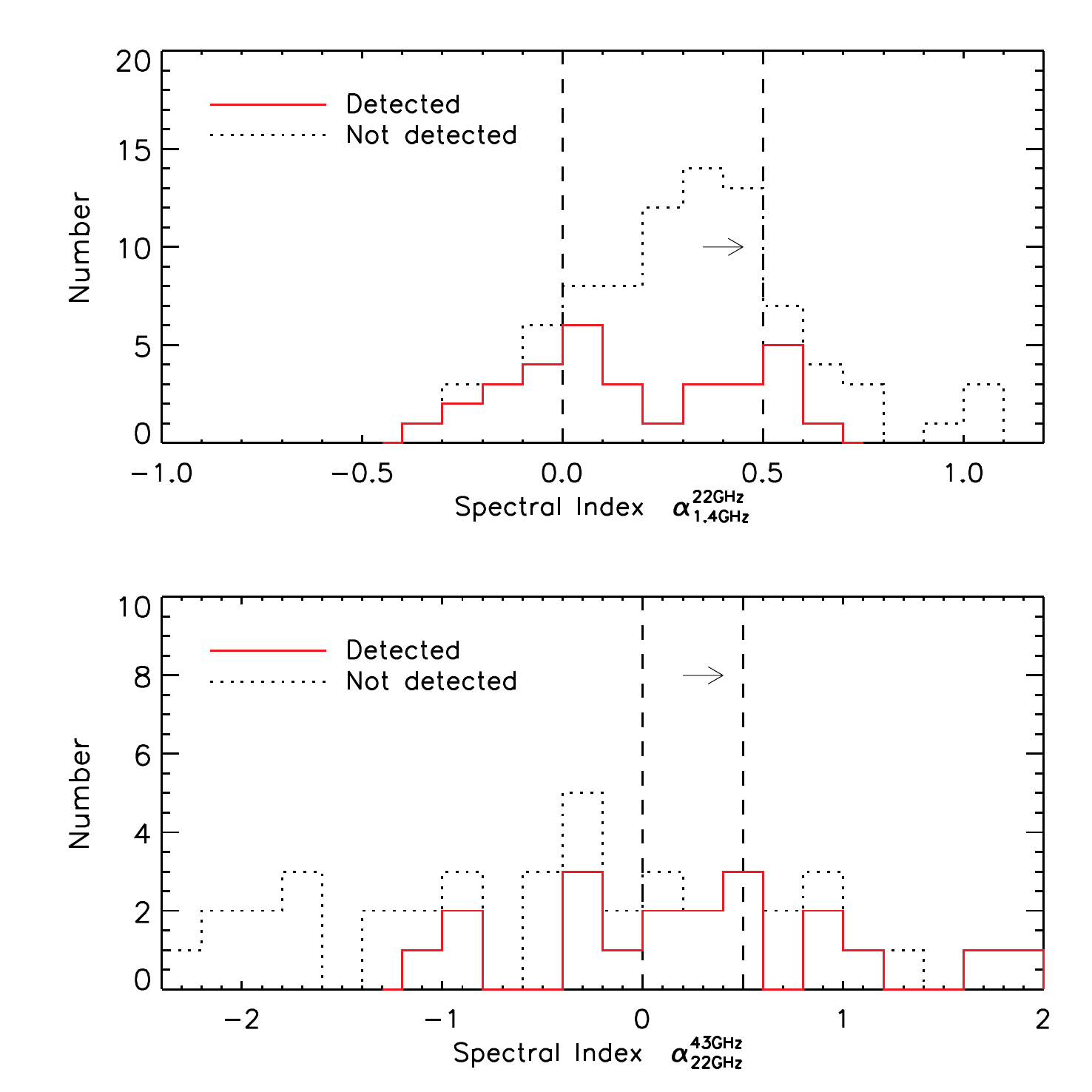}}
\caption[Spectral index distribution]
{Spectral index distribution. Distributions between 1.4 and 22 GHz (top) and between 22 and 43 GHz (bottom). The red solid line represents the detected galaxies at 22 (top) and at 43 GHz (bottom). Black dotted lines represent non-detected galaxies. The arrows in each panel indicate the lower limit of spectral indices for non-detected galaxies. Galaxies can be divided into three groups according to spectral index: steep ($\alpha > 0.5$), flat ($0 < \alpha < 0.5$), or inverted ($\alpha < 0$) spectra.}
\label{SpIdx}
\end{figure}

The distribution of spectral indices is shown in Fig.~\ref{SpIdx}. The top panel shows the distribution of the spectral index between 1.4 and 22 GHz. Out of the 32 galaxies detected at 22 GHz, 31\% (10/32) of the galaxies have inverted spectra, 50\% (16/32) have flat spectra, and 19\% (6/32) have steep spectra. The bottom panel shows the distribution of the spectral index between 22 and 43 GHz. Out of the 19 galaxies detected at 43 GHz, 7 (36\%), 6 (32\%), and 6 (32\%) galaxies have inverted, flat, and steep spectra, respectively. In both cases, most of the detected galaxies show either flat or inverted spectra rather than steep spectra. We also indicate non-detected galaxies (dotted line) based on the lower limit of the spectral index.

The spectral index enables us to infer optical depth at a given radio frequency. Radio emission from extended radio sources is optically thin with a steep spectrum. On the other hand, a flat or inverted spectrum is believed to be influenced by absorption mechanisms \citep{1973ranp.book.....P}. This implies that a number of detected galaxies in our sample might have compact cores at their centres. It appears that compact cores are common in AGNs. Several studies have revealed that compact cores with flat spectra occur in low luminosity AGNs \citep{1978Natur.276..374O, 2000ApJ...542..197F} and bright quasars \citep{2004A&A...418..429F}. Furthermore, flat-spectrum nuclear radio sources are found in both late-type galaxies \citep{2001ApJS..133...77H} and early-type galaxies \citep{1984ApJ...287...41W,  1991AJ....101..148W, 1989MNRAS.240..591S, 1994MNRAS.269..928S}.

\begin{figure}
\resizebox{\hsize}{!}{\includegraphics{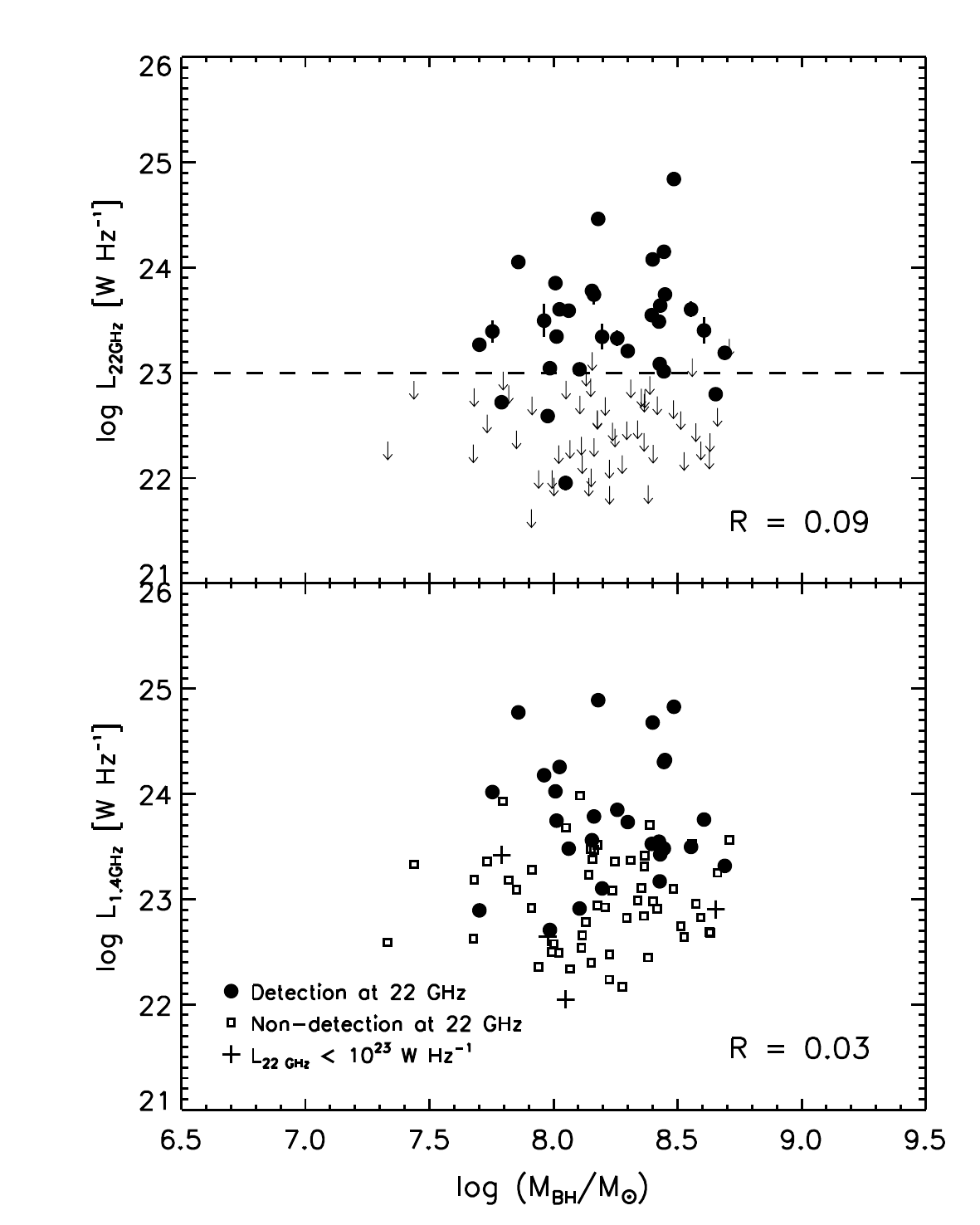}}
\caption{Relationship between radio luminosity and black hole mass for bright and detected galaxies at 22 GHz. Filled circles and open squares indicate detected and non-detected galaxies at 22 GHz. The crosses represent galaxies detected at 22 GHz with L$_{22GHz}$ $<$ 10$^{23}$ W Hz$^{-1}$. The correlation coefficients are derived from the sample of galaxies detected at 22 GHz L$_{22GHz}$ $>$ 10$^{23}$ W Hz$^{-1}$. There is no apparent correlation between black hole mass and radio luminosity at both 1.4 and 22 GHz.}
\label{blackhole_22}
\end{figure}

\begin{figure}
\resizebox{\hsize}{!}{\includegraphics{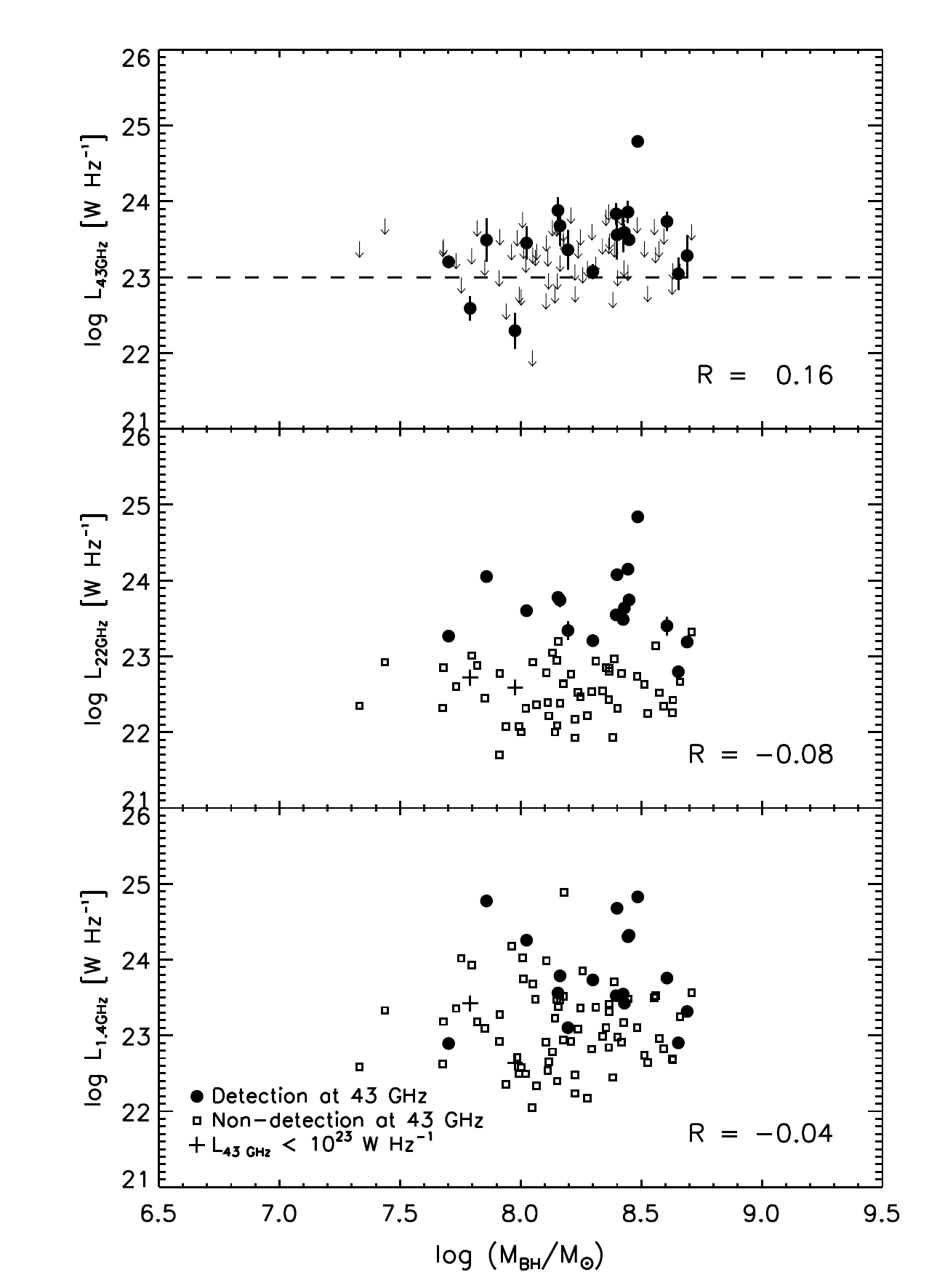}}
\caption[Relation between radio luminosity and black hole mass for bright and detected galaxies at 43 GHz]
{Relationship between radio luminosity and black hole mass for bright galaxies detected at 43 GHz. Filled circles and open squares indicate detected and non-detected galaxies at 43 GHz. The crosses represent detected galaxies with L$_{43GHz}$ $<$ 10$^{23}$ W Hz$^{-1}$. The correlation coefficients are derived from the sample of bright L$_{43GHz}$ $>$ 10$^{23}$ W Hz$^{-1}$ galaxies detected at 43 GHz. There is no apparent correlation between black hole mass and radio luminosity at any frequency.}
\label{blackhole_43}
\end{figure}

\subsection{Radio luminosity and black hole mass}
To derive the black hole mass of a galaxy, we used a well-known relation between black hole mass and velocity dispersion. The velocity dispersion $\sigma$ of a galaxy was extracted from the \cite{2011ApJS..195...13O} database. We used the formula from \cite{2006MNRAS.366.1126C} to correct for the difference in the apparent size of galaxies with respect to SDSS fibre coverage (1.5"),  
\begin{eqnarray}
\sigma_e = \sigma \left(\,\frac{R}{R_e}\,\right)^{0.066\,\pm\,0.035} ,
\end{eqnarray}
where $R$ is the aperture radius of a SDSS fibre, which is 1.5$''$, and $R_e$ is the effective circular radius of a galaxy. The SDSS catalogue provides an effective radius $R_{dev}$ derived from de Vaucouleurs model fitting to the surface brightness profile. We converted the effective radius to the effective circular radius $R_e$ by using the ratio of the isophotal major $IsoA_r$ and minor $IsoB_r$ axes in the SDSS $r$ band \citep{2003AJ....125.1817B}:
\begin{eqnarray}
R_e = \sqrt{\frac{IsoB_r}{IsoA_r}}\; R_{dev} .
\end{eqnarray}

We adopted the M$_{BH}$-$\sigma$ relation for elliptical galaxies from \cite{2009ApJ...698..198G}. Because we only considered elliptical galaxies in this study, Eq~(\ref{bhmass}) was applied to estimate the black hole mass of our sample galaxies:
\begin{eqnarray}
log\left( \frac{M_{BH}}{M_\odot}\right) = \left( 8.23 \pm 0.08\right) + \left( 3.96 \pm 0.42\right)log\left( \frac{\sigma_e}{200\,kms^{-1}}\right) .
\label{bhmass}
\end{eqnarray}
The black hole mass of our sample ranges from 10$^{7}$ to 10$^{9}$ M$_{\odot}$ and for detected galaxies at 22 GHz ranges from 10$^{7.7}$ to 10$^{8.7}$ M$_{\odot}$.

We investigated the relation between radio luminosities and the estimated black hole mass. In Fig.~\ref{blackhole_22}, radio power at 22 (top) and 1.4 GHz (bottom) is shown with respect to black hole mass. We evaluated the linear Pearson correlation coefficient R for bright galaxies, e.g. L$_{22GHz}$ $>$ 10$^{23}$ W Hz$^{-1}$, that are detected at 22 GHz. We then compared the correlation coefficients with respect to radio frequency in radio luminosity. No correlation is found between black hole mass and radio luminosity at either 1.4 or 22 GHz. 

Figure~\ref{blackhole_43} shows the relationship for detected and bright galaxies at 43 GHz. It does not show any trend in the coefficient. This suggests that radio luminosity is poorly related to black hole mass regardless of the observing frequencies in our sample.

The relationship between radio luminosity and black hole mass has been debated, with conflicting results. \cite{1998MNRAS.297..817F} found a tight correlation with a steep regression slope using a small number of nearby early-type galaxies. \cite{1999MNRAS.308..377M} showed a similar result for AGNs, but with a flatter regression slope. Since then, investigations based on larger quasar samples have reported the dependence of radio luminosity on black hole mass with a large scatter, concluding that radio loud quasars and radio galaxies have more massive black hole mass than radio quiet sources do \citep{2000ApJ...543L.111L, 2001ApJ...551L..17L, 2003MNRAS.340.1095D, 2004MNRAS.353L..45M, 2006MNRAS.365..101M}.

On the other hand, other studies have found no evidence for a relationship with larger samples ranging from inactive nearby galaxies to radio loud quasars \citep{2002ApJ...564..120H, 2002ApJ...576...81O, 2002ApJ...579..530W, 2005ApJ...631..762W}. \cite{2002ApJ...576...81O} and \cite{2002ApJ...579..530W} identified radio loud objects with black hole masses as low as 10$^6$ M$_{\odot}$. \cite{2003MNRAS.342..889S} investigated the relationship based on samples separated into inactive and active galaxies by using radio data at 1.4 GHz. They found a correlation of inactive galaxies similar to the \cite{1998MNRAS.297..817F} result, but no relationship for active galaxies. Our result shown in bottom panel of Figs.~\ref{blackhole_22} and~\ref{blackhole_43} is consistent with the findings of \cite{2003MNRAS.342..889S}. The previous studies mentioned above drew a conclusion using radio power at 1.4 GHz or 5 GHz, which are lower frequencies. Because compact regions are optically thin at high frequencies, we expect there to be a tighter correlation between black hole mass and radio luminosity at high frequencies if black hole mass is representative of AGN activity. In our sample, the relationship is poor. However, it should be remarked that black hole mass range of our sample is rather narrow, namely 10$^{7.7}$ to 10$^{8.7}$ M$_{\odot}$.

\begin{figure}
\resizebox*{\hsize}{!}{\includegraphics{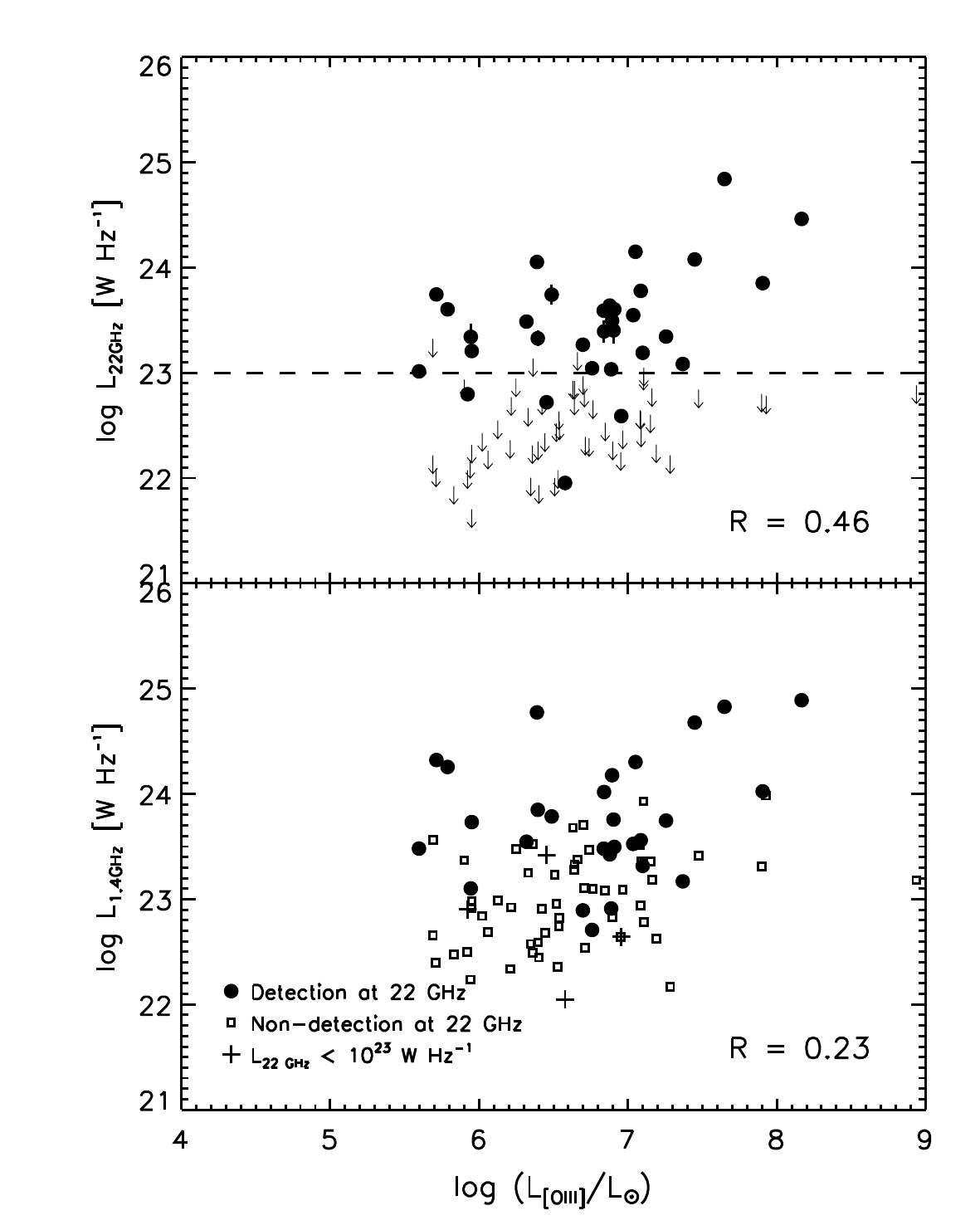}}
\caption[Relation between radio luminosity and {[OIII]} emission line luminosity for bright and detected galaxies at 22 GHz]
{Relation between radio luminosity and [OIII] emission line luminosity for bright and detected galaxies at 22 GHz. Filled circles and open squares indicate the detected and non-detected galaxies at 22 GHz. The crosses represent the detected galaxies, but with L$_{22GHz}$ $<$ 10$^{23}$ W Hz$^{-1}$. The correlation coefficients are derived from the sample of bright L$_{22GHz}$ $>$ 10$^{23}$ W Hz$^{-1}$ and detected galaxies at 22 GHz.}
\label{lumO22}
\end{figure}

\begin{figure}
\resizebox*{\hsize}{!}{\includegraphics{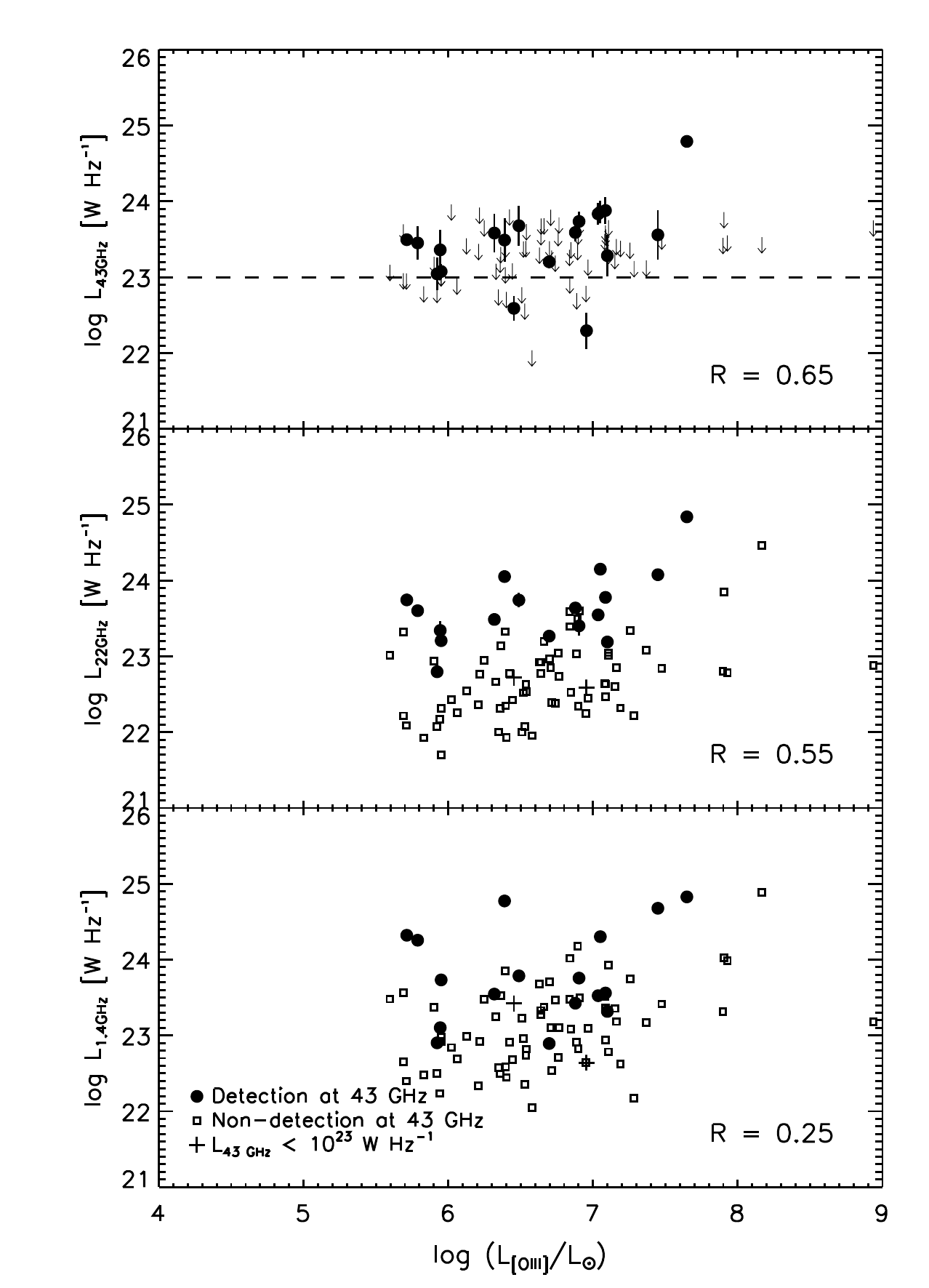}}
\caption[Relation between radio luminosity and {[OIII]} emission line luminosity for bright and detected galaxies at 43 GHz]
{Relation between radio luminosity and [OIII] emission line luminosity for bright and detected galaxies at 43 GHz. Filled circles and open squares indicate the detected and non-detected galaxies at 43 GHz. The crosses represent the detected galaxies, but with L$_{43GHz}$ $<$ 10$^{23}$ W Hz$^{-1}$. The correlation coefficients are derived from the sample of bright L$_{43GHz}$ $>$ 10$^{23}$ W Hz$^{-1}$ and detected galaxies at 43 GHz.}
\label{lumO43}
\end{figure}

\subsection{Radio luminosity and [OIII] emission line luminosity}
There is a relationship between the radio power and luminosity of optical emission lines in active galaxies such as Seyfert \citep{1978A&A....64..433D, 1985MNRAS.213...33W, 1992ApJS...79...49W, 1996ApJ...468..475G} and powerful radio galaxies \citep{1989ApJ...336..702B, 2011MNRAS.415.1013K}. 
Ka03 and \cite{2004ApJ...613..109H} suggested that the luminosity of the [OIII] $\lambda$5007 emission line can be used as a tracer of AGN activity. This is supported by the fact that this forbidden line is one of the strongest lines in optical spectra, and is seen in the narrow-line region (NLR) of AGN. Moreover, the [OIII] emission line suffers very little contamination due to star forming activity within a host galaxy. \cite{2012MNRAS.422.3268J} found a strong correlation between the [OIII] luminosity and hard X-ray luminosity in Type I AGNs, which suggests that the [OIII] emission line can be used to estimate the strength of AGN activity \citep{1994ApJ...436..586M}.

We discussed the relationship between radio and [OIII] $\lambda$5007 emission line luminosity and investigated whether radio emission by AGN is relevant to optical AGN activity. In Fig.~\ref{lumO22}, we show the linear Pearson correlation coefficient R of galaxies detected at 22 GHz, L$_{22GHz}$ $>$ 10$^{23}$ W Hz$^{-1}$. Unlike the relationship between radio power and black hole mass shown in Figs.~\ref{blackhole_22} and~\ref{blackhole_43}, the correlation coefficient increases with the observation frequency. Fig.~\ref{lumO43} shows the result for galaxies detected at 43 GHz. This result shows a trend similar to that illustrated in Fig.~\ref{lumO22}. The relationship becomes tighter as the radio frequency increases. 
These results show that there is a tighter correlation at high frequencies compared to 1.4 GHz. This confirms that observations at high frequencies can reveal compact central regions that are too opaque to be seen at low frequencies. This also suggests that the [OIII] emission line can be used as an indicator of AGN activity, and that radio and optical AGN activity are related and may originate from the same mechanism.

\cite{2005MNRAS.362...25B} found that [OIII] luminosity is not associated with radio luminosity at 1.4 GHz using NVSS survey data. This coincides well with our result at 1.4 GHz, as shown in the bottom panel of Figs.~\ref{lumO22} and~\ref{lumO43}. They suggested that radio and optical AGN activity are the results of different mechanisms. However, their sample of radio loud AGNs could be affected by absorption mechanisms. On the other hand, \cite{2000ApJ...542..186N} found that radio power at 15 GHz correlates with the luminosity of the [OI] $\lambda$6300 emission line for low luminosity AGNs (LLAGNs). \cite{2006A&A...447...97B} showed that the low luminosity radio galaxies (LLRGs) in their sample are genuine active nuclei that host a radio loud core. They found correlations between X-ray, optical, and radio luminosities at 5 GHz, suggesting a common non-thermal origin of nuclear emission. 

This discrepancy can be explained in terms of radio opacity at different frequencies. In conclusion, assuming that the luminosity of the [OIII] $\lambda$5007 emission line traces AGN activity, radio luminosity at a high radio frequency can be used to test the dependence of radio power and optical AGN activity.

\begin{figure}
\resizebox{\hsize}{!}{\includegraphics{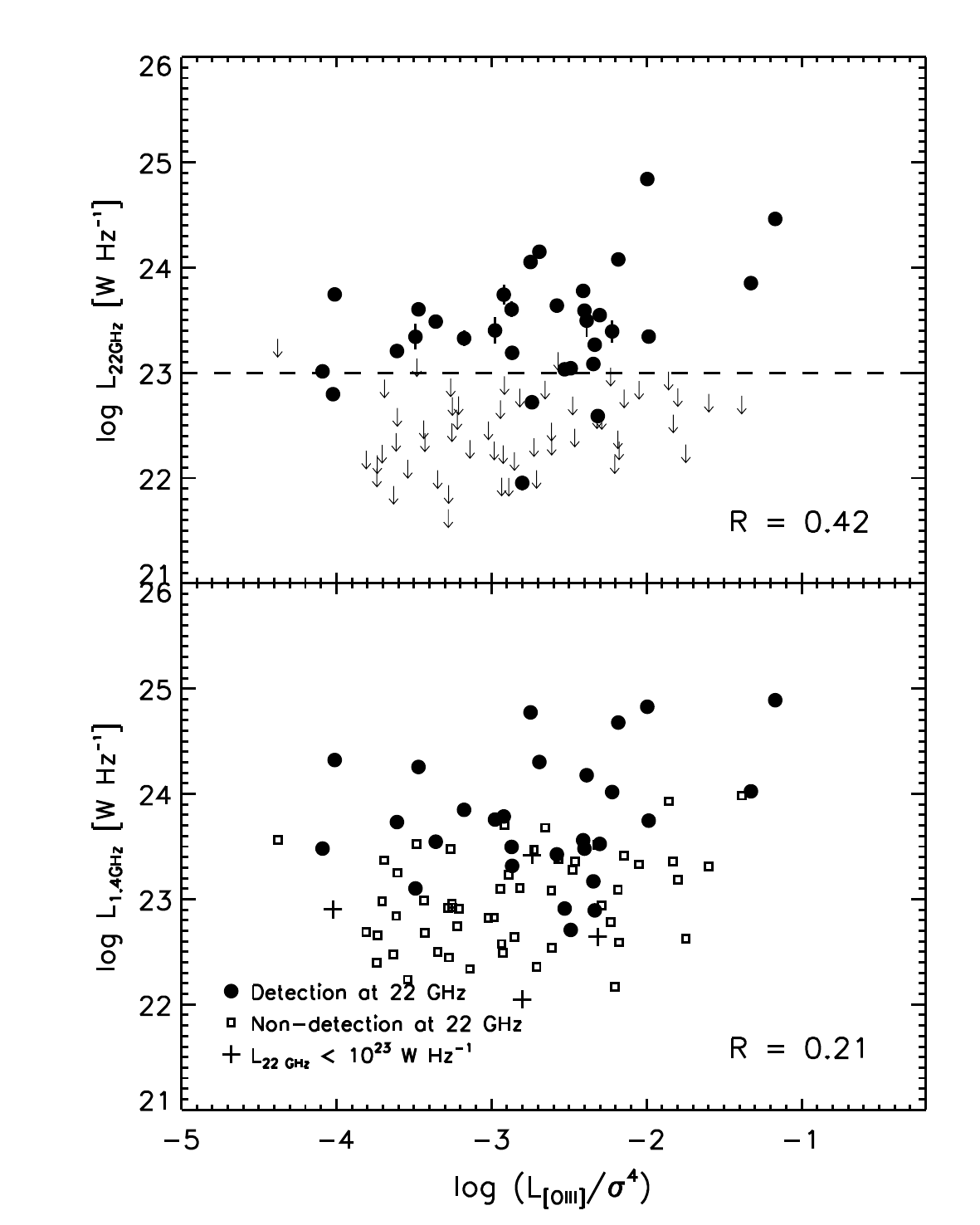}}
\caption[Relation between radio luminosity and the Eddington ratio for bright and detected galaxies at 22 GHz]
{Relation between radio luminosity and the Eddington ratio for bright and detected galaxies at 22 GHz. Filled circles and open squares indicate the detected and non-detected galaxies at 22 GHz. The crosses represent the detected galaxies, but with L$_{22GHz}$ $<$ 10$^{23}$ W Hz$^{-1}$. The correlation coefficients are derived from the sample of bright L$_{22GHz}$ $>$ 10$^{23}$ W Hz$^{-1}$ and detected galaxies at 22 GHz.}
\label{lumE22}
\end{figure}

\begin{figure}
\resizebox{\hsize}{!}{\includegraphics{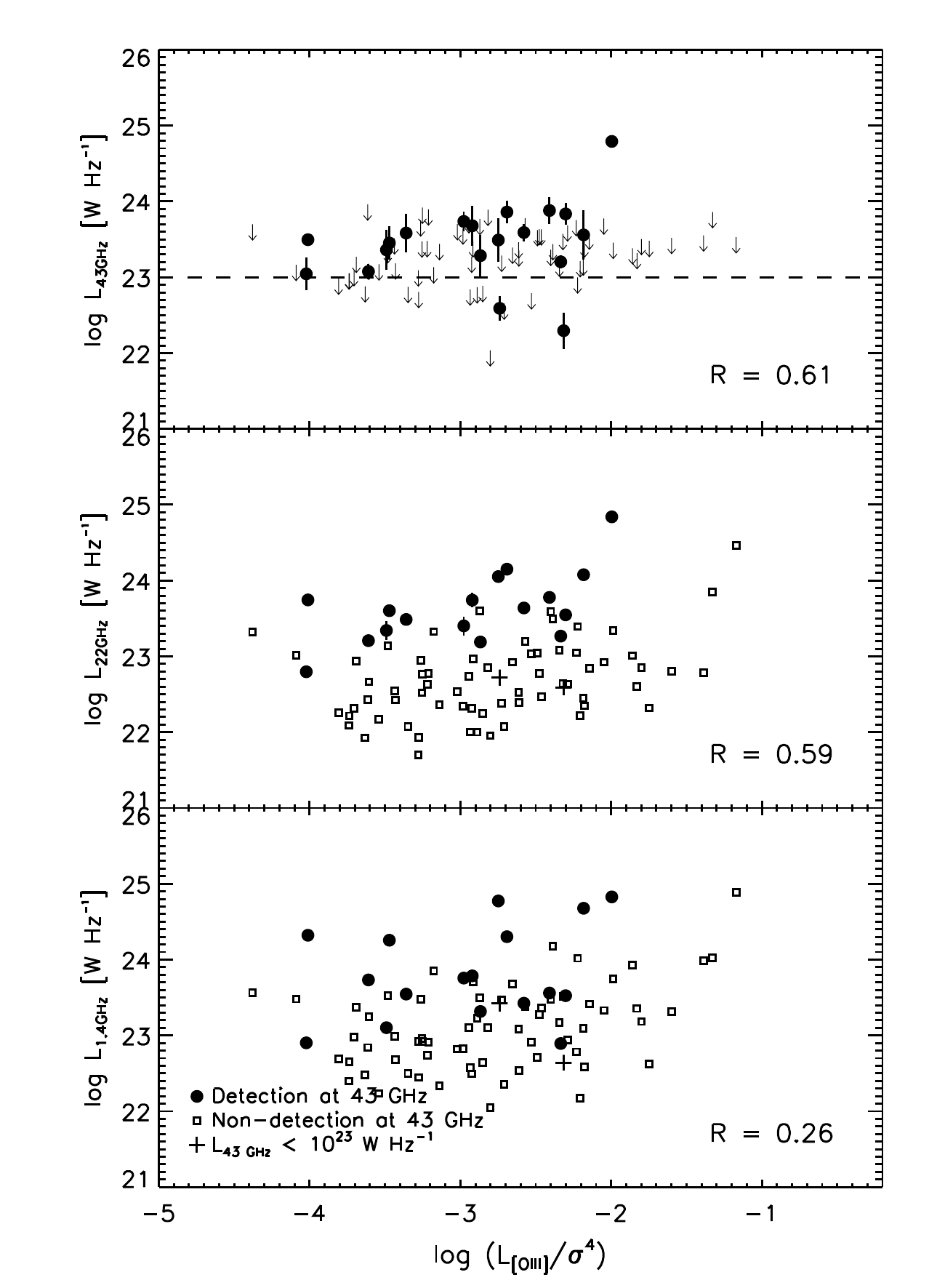}}
\caption[Relation between radio luminosity and Eddington ratio for bright and detected galaxies at 43 GHz]
{Relation between radio luminosity and Eddington ratio for bright and detected galaxies at 43 GHz. Filled circles and open squares indicate the detected and non-detected galaxies at 43 GHz. The crosses represent the detected galaxies, but with L$_{43GHz}$ $<$ 10$^{23}$ W Hz$^{-1}$. The correlation coefficients are derived from the sample of bright L$_{43GHz}$ $>$ 10$^{23}$ W Hz$^{-1}$ and detected galaxies at 43 GHz.}
\label{lumE43}
\end{figure}

\subsection{Radio luminosity and the Eddington ratio}
In this section, we investigate the relationship between radio luminosity and mass accretion rate on black holes. The Eddington ratio defined as L$_{bol}$/L$_{Edd}$ can be used as a proxy for the accretion rate \citep{2009MNRAS.397..135K}, where L$_{bol}$ is the bolometric luminosity and L$_{Edd}$ is the Eddington luminosity. The [OIII] luminosity can be used as an indicator of the bolometric luminosity of the central black hole \citep{2004ApJ...613..109H, 2008MNRAS.385.1915L}. The Eddington luminosity is proportional to the black hole mass, which can be estimated from stellar velocity dispersion. Consequently, the Eddington ratio can be represented by L$_{[OIII]}/\sigma^{4}$. 

Figures~\ref{lumE22} and~\ref{lumE43} show the radio luminosities versus the Eddington ratio. The correlation coefficient is derived from galaxies detected at 22 GHz. Likewise, we obtained the correlation coefficient using galaxies detected at 43 GHz. The correlation with the Eddington ratio increases with the frequency of radio luminosity which shows a trend similar to the result of the relationship with the [OIII] emission line luminosity.

\cite{2012MNRAS.421.1569B} suggested that the dichotomy in radio AGN is attributed mainly to the difference in accretion rate. Based on their classification scheme, our sample corresponds to low excitation radio galaxies, where L$_{1.4GHz} < 10^{26}$ W Hz$^{-1}$. They concluded that it is difficult to find the correlation in low excitation radio galaxies associated with low black hole accretion rates. Nevertheless, galaxies with high Eddington ratios ranging from 0.1 to 0.01 do exist, which suggests that these galaxies are influenced by other factors, such as black hole spin.

\begin{figure}
\resizebox{\hsize}{!}{\includegraphics{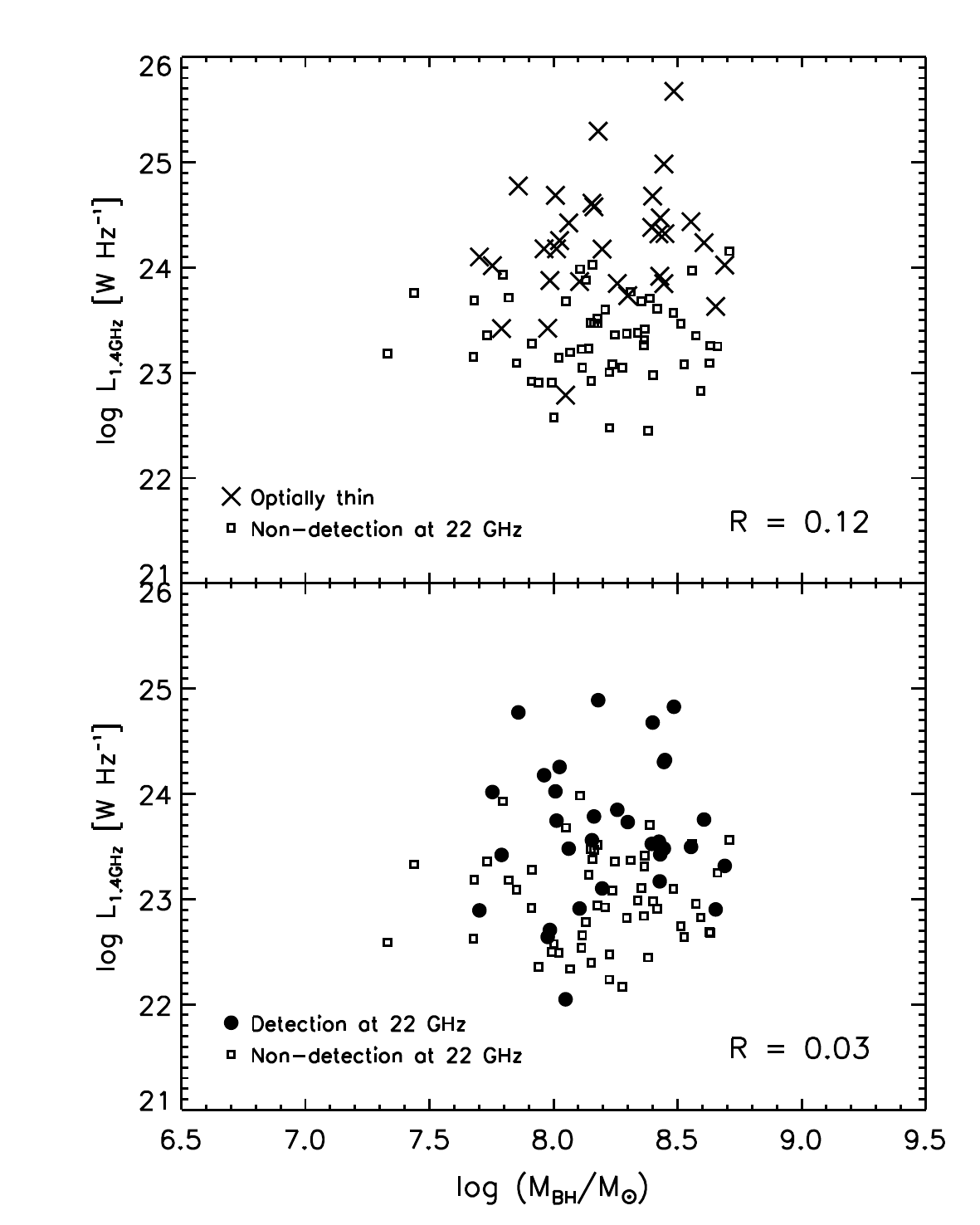}}
\caption[Radio luminosity at 1.4 GHz and black hole mass]
{Radio luminosity at 1.4 GHz and black hole mass. Filled circles and open squares indicate the detected and non-detected galaxies at 22 GHz. The Xs represent luminosity at 1.4 GHz extrapolated from the 22 GHz flux assuming steep power law at 1.4 GHz by estimation from 22 GHz. It shows a similar result indicating independence on black hole mass.} 
\label{optthinbh}
\end{figure}

\begin{figure}
\resizebox{\hsize}{!}{\includegraphics{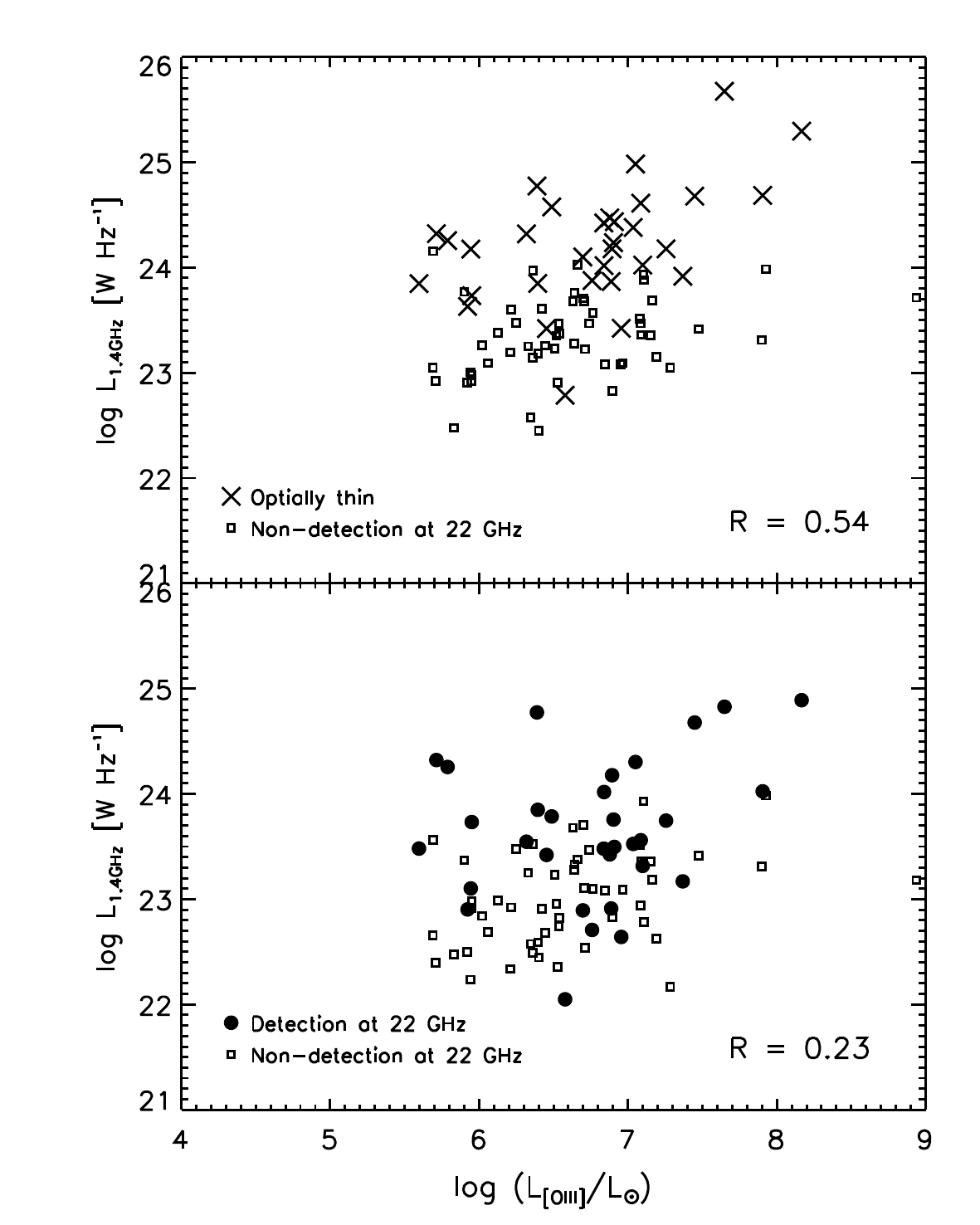}}
\caption[Radio luminosity at 1.4 GHz and {[OIII]} emission line luminosity]
{Radio luminosity at 1.4 GHz and [OIII] emission line luminosity. Filled circles and open squares indicate the detected and non-detected galaxies at 22 GHz. The Xs represent luminosity at 1.4 GHz extrapolated from the 22 GHz flux assuming steep power law at 1.4 GHz by estimation from 22 GHz. It suggests that optically thin conditions make it better to trace AGN activity.}
\label{optthinemission}
\end{figure}

\begin{figure}
\resizebox{\hsize}{!}{\includegraphics{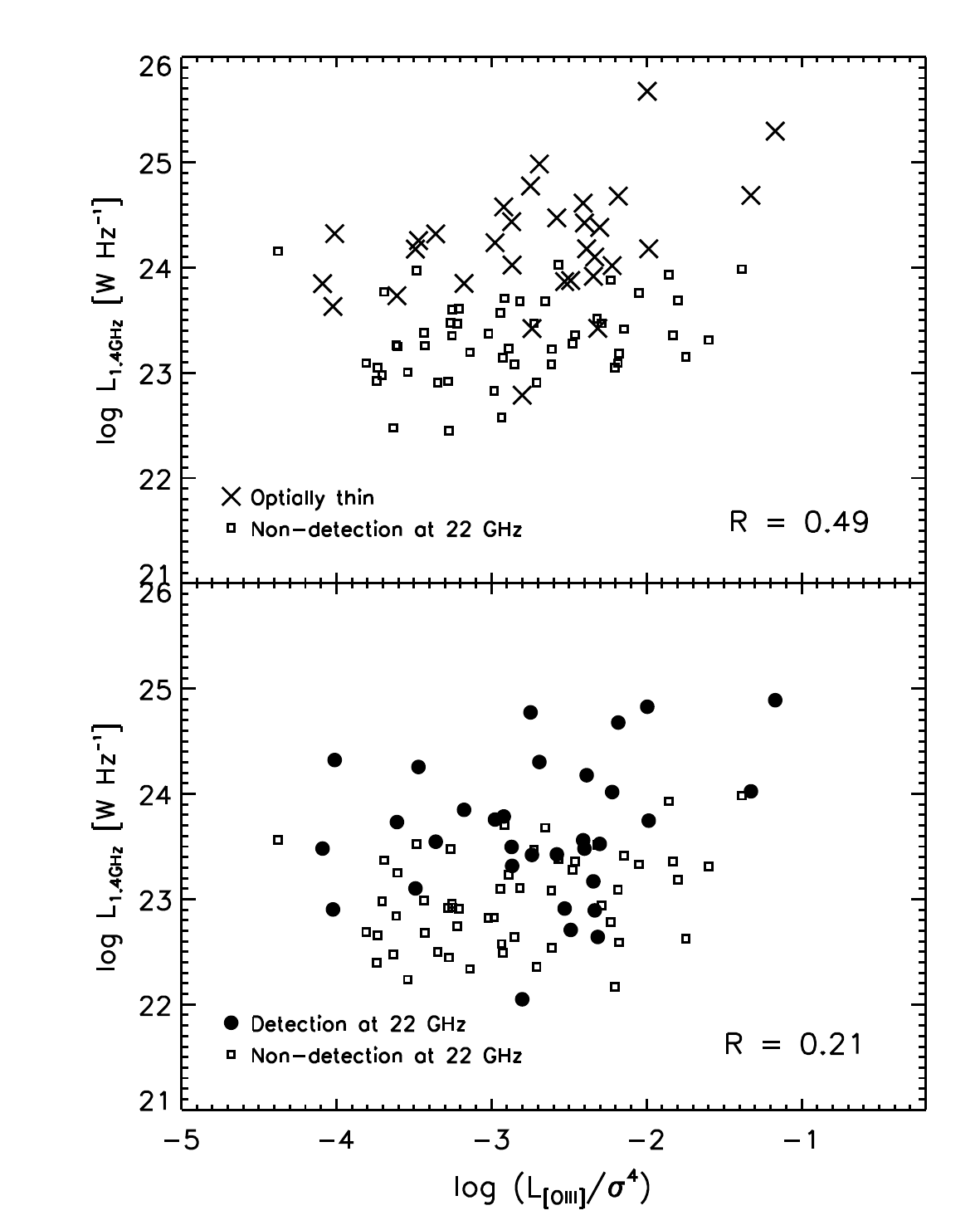}}
\caption[Radio luminosity at 1.4 GHz and Eddington ratio]
{Radio luminosity at 1.4 GHz and Eddington ratio. Filled circles and open squares indicate the detected and non-detected galaxies at 22 GHz. The Xs represent luminosity at 1.4 GHz extrapolated from the 22 GHz flux assuming steep power law at 1.4 GHz by estimation from 22 GHz. Similar to Figure~\ref{optthinemission}, it shows a tighter correlation in optically thin conditions.}
\label{optthinedd}
\end{figure}

\subsection{Inferred radio power at 1.4 GHz}
We confirmed that radio power at high radio frequencies is more likely to correlate with the properties of AGNs, with the exception of black hole mass. Radio power at 1.4 GHz can be underestimated because of synchrotron self-absorption or free-free absorption. To compare the observed radio power at 1.4 GHz with the estimation and to confirm the effects of these absorptions, we revisited the relationship between radio power and the properties of AGNs.

In order to estimate radio power at 1.4 GHz, we derived optically thin radio luminosity at 1.4 GHz, assuming that synchrotron radiation is optically thin with $\alpha \sim$ 0.5$-$0.7. Because of the steeper slope of synchrotron radiation, the estimated radio power at 1.4 GHz is greater than the observed radio power by the FIRST survey. Correlation coefficients were derived from the sample of galaxies detected at 22 GHz, L$_{22GHz}$ $>$ 10$^{23}$ W Hz$^{-1}$, in the same manner as in previous sections.

Figure~\ref{optthinbh} shows radio luminosity at 1.4 GHz versus black hole mass. The correlation coefficients indicate that radio power does not correlate with black hole mass and is thus not a good indicator of black hole mass regardless of opacity. This is consistent with the results from Sect. 4.3. Figure~\ref{optthinemission} shows the relationship between radio power and the luminosity of the [OIII] emission line. Similarly, Fig.~\ref{optthinedd} shows the relationship between radio power and the Eddington ratio. In both cases, we found that radio power derived from an optically thin condition correlates with optical AGN activity. These results are in agreements with the conclusion of previous sections examining radio power at 22 or 43 GHz and the properties of AGNs.

\section{Discussion}
We performed radio continuum observations at 22 and 43 GHz, which are an order of magnitude higher than the 1.4 or 5 GHz widely used for surveys and statistical studies. We revisited the relationship between radio luminosities and the properties of AGNs, such as $\sigma^4$, [OIII] emission line luminosity, and the Eddington ratio.

The correlation between radio power and AGN properties becomes tighter at higher frequencies. This implies that the correlation is obscured at low frequencies, since radio emission is optically thick. This coincides well with our results of the distribution of the spectral index. In our sample, 81\% (68\%) of sources have a flat or inverted spectrum between 1.4 and 22 GHz (22 and 43 GHz). There is no significant relationship between radio luminosity and black hole mass, although it should be noted that the black hole mass range is rather narrow (10$^{7.7-8.7}$ M$_{\odot}$ for 22 GHz detection). On the other hand, the indicator of mass accretion L$_{[OIII]}/\sigma^{4}$ has a range from 10$^{-4.1}$ to 10$^{-1.1}$. In our sample, radio luminosity correlates with [OIII] emission line luminosity and the Eddington ratio. 

Statistical studies have used radio loud AGNs including beaming sources e.g. blazars. Their radio flux is Doppler boosted. \cite{2002MNRAS.336L..38J} re-examined the relationship for the same sample as \cite{2002ApJ...576...81O}, who had found the relationship between radio luminosity and black hole mass to be independent. They pointed out that Doppler boosting may significantly influence the relationship and claimed that the correlation becomes significant after correcting for this effect. Since we used Type II AGNs, which have edge-on accretion discs and thick dusty tori according to the AGN unification model \citep{1995PASP..107..803U}, our sample is not strongly affected by Doppler boosting. Nonetheless, a correlation was not found in our sample. This could be mainly because present-day black hole mass represents integrated mass accretion, whereas radio emission is derived from episodic AGN activity on a timescale of $\sim$ 10$^{7-8}$ years \citep{2007MNRAS.382.1019B, 2011MNRAS.417L..36H}. Even though a marginal threshold of black hole mass is needed to turn on AGN radio activity \citep{2006MNRAS.365..101M}, once radio activity is triggered, it seems that radio luminosity is not dependent on black hole mass. We should point out that the range of black hole mass in our sample is not sufficiently wide to see the dependence on radio luminosity. In order to identify a principal connection with radio power, we plan to investigate this relationship on a larger range of black hole masses and at a certain mass accretion rate.

The connection between radio power and optical emission line strength has been discussed in the context of quasars and powerful radio galaxies. Among the strong optical emission lines, we used the [OIII] $\lambda$5007 emission line. The simplest interpretation is that the accretion rate directly influences the fuelling of radio jet power, as suggested by some models \citep[i.e.][]{1995A&A...293..665F, 1999AJ....118.1169X}. \cite{2001ApJ...555..650H} argued that the radio power traces the accretion luminosity. In the large samples used in that study, the range of radio luminosity for a given black hole mass is wide, and is probably more dependent on the black hole accretion rate than on mass \citep{2002ApJ...564..120H}. \cite{2002ApJ...564..120H} suggested that the relationship between radio luminosity and black hole mass arises indirectly through more fundamental correlations between radio luminosity and bulge mass and between bulge mass and black hole mass. This argument is similar to the original interpretation of \cite{1996ApJ...465...96N}. 

The correlation between optical lines and the radio power of AGNs could lead us to an incorrect interpretation when we take the radio power from extended radio lobes or at low frequency from compact regions. It has been reported that the galaxy in our sample, 1140+1743 (NGC 3801), does not show radio core at 1.4 GHz but at 100 GHz \citep{2005ApJ...629..757D}. This source is detected at both 22 and 43 GHz and has a flat spectra between 22 and 43 GHz. We cannot distinguish core flux from total flux with single-dish observation. However, the flux we measured can be regarded as core flux because radio emission at 22 or 43 GHz is believed to originate from the compact cores at the centres of galaxies. The detection in our observation can support the existence of radio compact cores.

As seen in NGC 3801 and in the episodic radio galaxy Speca, the central optical emission lines reflect the nature of the source that need not have originated at the epoch of radio-jet ejection, visible to the radio observations \citep{2011MNRAS.417L..36H, 2012MNRAS.422L..38H}. Hence, radio-optical comparisons need to be carefully interpreted. Recently the relics of AGN activity have been reported \citep{2012MNRAS.420..878K, 2013ApJ...763...60S} and their ages seem to be young with a time scale of 10$^{3-4}$ years. It is interesting to note that radio sources like the gigahertz peaked spectrum (GPS) or high frequency peakers (HFP) have ages of thousand years. Some of detected galaxies in our observations can be GPS or HFP candidates.

In this paper, we showed the evidence of correlation between radio power and accretion rate of AGNs. In order to study whether the black hole mass relates to radio power, we will explore wider black hole mass range samples. We are also going to perform high radio frequency VLBI observations to investigate the optically thick compact radio structures. This may provide new insight into the co-evolution of AGNs and their host galaxies in the era of EVLA and ALMA.

\begin{acknowledgements}
We thank the anonymous referee for helpful comments and suggestions. This paper is a result of the collaborative project between Yonsei University and Korea Astronomy and Space Science Institute through DRC program of Korea Research Council of Fundamental Science and Technology(FY 2012). We also acknowledge support from the National Research Foundation of Korea through Doyak Grant (No. 20090078756), the Center for Galaxy Evolution Research Grant (No. 2010-0027910), and S.Y.P. is grateful for support by the International Research \& Development Programm of the National Research Foundation of Korea (NRF) funded by the Ministry of Education, Science and Technology (MEST) of Korea (Grant number: 2012049606, FY2012). S.K.Y. and B.W.S. both acted as corresponding authors. 
\end{acknowledgements}

\bibliographystyle{aa} 
\bibliography{PSY} 

\tiny
\begin{longtab}
\begin{longtable}{c c c c c c c c c c c c c}
\caption[KVN single-dish observations results]{KVN single-dish observation results} \label{tab:KVN_detection} \\
\hline \hline
\multirow{2}{*}{Source} & \multirow{2}{*}{RA} & \multirow{2}{*}{DEC} & \multirow{2}{*}{Date} & $f_{1.4GHz}$\tablefootmark{a} & $\sigma_{1.4GHz}$\tablefootmark{a} & $f_{22GHz}$ & $\sigma_{22GHz}$ & $f_{43GHz}$ & $\sigma_{43GHz}$ &  \multirow{2}{*}{$\alpha_{1.4GHz}^{22GHz}$} & \multirow{2}{*}{$\alpha_{22GHz}^{43GHz}$} \\
& & & & (mJy)  & (mJy)  & (mJy)  & (mJy)  & (mJy) & (mJy) & \\
\hline
\endfirsthead
\caption[]{KVN single-dish observation results (continued)} \\
\hline \hline
\multirow{2}{*}{Source} & \multirow{2}{*}{RA} & \multirow{2}{*}{DEC} & \multirow{2}{*}{Date} & ${f_{1.4GHz}}$\tablefootmark{a} & ${\sigma_{1.4GHz}}$\tablefootmark{a} & $f_{22GHz}$ & $\sigma_{22GHz}$ & $f_{43GHz}$ & $\sigma_{43GHz}$ & \multirow{2}{*}{$\alpha_{1.4GHz}^{22GHz}$} & \multirow{2}{*}{$\alpha_{22GHz}^{43GHz}$} \\
& & & & (mJy)  & (mJy)  & (mJy)  & (mJy)  & (mJy) & (mJy) & \\
\hline
\endhead
\hline
\endfoot
\hline \hline
\endlastfoot
 0041$-$0918& 00:41:50.5  &  $-$09:18:11.3  & 2010 Nov. 28 &  53.2 (40.2) &    0.2  & ($<$30.4)  &     -  &   ($<$72.5)  &    -  &    ($>$0.2)  &   ($>$-1.3) \\
 0225$-$0841&     02:25:16.9  &  $-$08:41:37.5  & 2011 Jan. 01 &  23.6  &    0.2  &   ($<$11.1)  &       -  &   ($<$49.6)  &       -  &    ($>$0.3)  &   ($>$-2.2)   \\
 0752+4556  &     07:52:44.2  &  $+$45:56:57.4  & 2010 Feb. 14 & 175.8 (48.6)  &    0.2  &       118.1  &     2.5  &  ($<$120.5)  &       -  &         0.1  &   ($>$-0.0)   \\
 0756+5116  &     07:56:26.3  &  $+$51:16:33.7  & 2010 Dec. 28 &  32.8  &    0.1  &    ($<$4.2)  &       -  &   ($<$61.4)  &       -  &    ($>$0.7)  &   ($>$-4.0)   \\
 0758+3747  &     07:58:28.1  &  $+$37:47:11.7  & 2010 Nov. 27 & 546.0 (25.8) &    0.3 &       383.8  &     8.7  &       196.9  &    29.8  &         0.1  &         1.0   \\
 0805+2409  &     08:05:35.0  &  $+$24:09:50.4  & 2011 Feb. 15 & 960.5 (-) &    0.6 &       357.9  &     5.9  &   ($<$41.9)  &       -  &         0.4  &    ($>$3.2)   \\
 0831+3535  &     08:31:19.3  &  $+$35:35:48.8  & 2011 Feb. 19 &  18.8  &    0.2  &   ($<$10.4)  &       -  &  ($<$115.1)  &       -  &    ($>$0.2)  &   ($>$-3.6)   \\
 0836+4401  &     08:36:37.8  &  $+$44:01:09.6  & 2011 Jan. 02 & 139.3  &    0.1  &    ($<$8.8)  &       -  &   ($<$51.6)  &       -  &    ($>$1.0)  &   ($>$-2.6)   \\
 0901+1037  &     09:01:00.1  &  $+$10:37:01.8  & 2011 Mar. 26 &  50.4 (44.9) &    0.1  &   ($<$10.9)  &       -  &   ($<$65.9)  &       -  &    ($>$0.6)  &   ($>$-2.7)   \\
 0904+4219  &     09:04:43.4  &  $+$42:19:13.2  & 2011 Feb. 22 &  10.7  &    0.2  &    ($<$9.2)  &       -  &   ($<$92.4)  &       -  &    ($>$0.1)  &   ($>$-3.4)   \\
 0910+1841  &     09:10:39.9  &  $+$18:41:47.7  & 2010 Dec. 30 &  47.4  &    0.1  &    ($<$2.9)  &       -  &   ($<$70.5)  &       -  &    ($>$1.0)  &   ($>$-4.8)   \\
 0913+2959  &     09:13:39.5  &  $+$29:59:34.7  & 2011 Jan. 01 &  25.6  &    0.1  &    ($<$7.8)  &       -  &   ($<$58.5)  &       -  &    ($>$0.4)  &   ($>$-3.0)   \\
 0930+3413  &     09:30:03.6  &  $+$34:13:25.3  & 2010 Dec. 31 &  30.8  &    0.1  &    ($<$8.5)  &       -  &   ($<$73.6)  &       -  &    ($>$0.5)  &   ($>$-3.2)   \\
 0933+1009  &     09:33:46.1  &  $+$10:09:09.0  & 2010 Dec. 28 &  45.6  &    0.2  &        36.6  &     2.5  &   ($<$44.4)  &       -  &         0.1  &   ($>$-0.3)   \\
 0943+3614  &     09:43:19.2  &  $+$36:14:52.2  & 2011 Jan. 02 &  74.9  &    0.1  &       176.9  &     2.2  &       152.9  &    12.6  &        -0.3  &         0.2   \\
 0954+1036  &     09:54:02.2  &  $+$10:36:29.7  & 2011 Feb. 21 &  18.9  &    0.1  &    ($<$9.8)  &       -  &  ($<$143.5)  &       -  &    ($>$0.2)  &   ($>$-4.0)   \\
 1020+4831  &     10:20:53.7  &  $+$48:31:24.1  & 2010 Dec. 31 &  75.0  &    0.1  &   ($<$13.2)  &       -  &   ($<$38.8)  &       -  &    ($>$0.6)  &   ($>$-1.6)   \\
 1025+1022  &     10:25:44.2  &  $+$10:22:30.5  & 2011 Mar. 23 &  75.7  &    0.1  &        66.0  &     2.2  &        82.6  &    21.0  &         0.0  &        -0.3   \\
 1032+5644  &     10:32:58.9  &  $+$56:44:53.4  & 2010 Nov. 26 & 127.9 (29.4) &    0.2 &        56.5  &     7.1  &       122.1  &    15.6  &         0.3  &        -1.1   \\
 1037+4335  &     10:37:19.3  &  $+$43:35:15.3  & 2011 Jan. 02 & 128.9  &    0.1  &    ($<$7.6)  &       -  &   ($<$56.0)  &       -  &    ($>$1.0)  &   ($>$-3.0)   \\
 1043+3131  &     10:43:18.6  &  $+$31:31:06.2  & 2010 Nov. 26 & 648.4 (401.7) &    0.1 &       144.0  &     7.6  &       101.9  &    22.4  &         0.5  &         0.5   \\
 1044+1117  &     10:44:35.8  &  $+$11:17:46.1  & 2011 Feb. 19 &  19.7  &    0.1  &    ($<$8.6)  &       -  &   ($<$96.7)  &       -  &    ($>$0.3)  &   ($>$-3.6)   \\
 1044+4354  &     10:44:03.7  &  $+$43:54:12.1  & 2011 Jan. 02 &  23.2  &    0.1  &    ($<$8.7)  &       -  &   ($<$55.2)  &       -  &    ($>$0.4)  &   ($>$-2.8)   \\
 1048+0459  &     10:48:11.9  &  $+$04:59:54.9  & 2010 Dec. 31 &  48.0  &    0.2  &   ($<$11.0)  &       -  &   ($<$66.4)  &       -  &    ($>$0.5)  &   ($>$-2.7)   \\
 1059+0517  &     10:59:14.6  &  $+$05:17:31.3  & 2010 Nov. 26 &  46.2 (34.7) &    0.2 &        80.2  &     9.9  &        84.0  &    22.4  &        -0.2  &        -0.1   \\
 1107+5552  &     11:07:59.9  &  $+$55:52:50.8  & 2010 Feb. 18 &  10.6  &    0.1  &   ($<$19.5)  &       -  &   ($<$99.7)  &       -  &   ($>$-0.2)  &   ($>$-2.4)   \\
 1111+2657  &     11:11:25.2  &  $+$26:57:49.0  & 2010 Nov. 27 &  72.4 (12.8) &    0.1  &   ($<$18.8)  &       -  &   ($<$61.5)  &       -  &    ($>$0.5)  &   ($>$-1.8)   \\
 1116+2915  &     11:16:22.7  &  $+$29:15:08.3  & 2011 Mar. 22 &  73.1  &    0.2  &   ($<$30.1)  &       -  &   ($<$55.0)  &       -  &    ($>$0.3)  &   ($>$-0.9)   \\
 1117+3235  &     11:17:00.1  &  $+$32:35:51.0  & 2011 Feb. 19 &  19.2  &    0.1  &        41.6  &     2.8  &  ($<$158.0)  &       -  &        -0.3  &   ($>$-2.0)   \\
 1122+3406  &     11:22:56.5  &  $+$34:06:41.4  & 2011 Feb. 20 &  16.6  &    0.1  &    ($<$5.5)  &       -  &  ($<$109.3)  &       -  &    ($>$0.4)  &   ($>$-4.5)   \\
 1127+4004  &     11:27:27.5  &  $+$40:04:09.5  & 2011 Feb. 22 &  14.8  &    0.1  &    ($<$8.5)  &       -  &  ($<$113.1)  &       -  &    ($>$0.2)  &   ($>$-3.9)   \\
 1140+1743  &     11:40:17.0  &  $+$17:43:40.4  & 2010 Nov. 29 & 918.2 (-) &    0.1 &       183.0  &     6.8  &       135.7  &    22.7  &         0.6  &         0.4   \\
 1145+1936  &     11:45:05.0  &  $+$19:36:22.8  & 2010 Nov. 28 & 2090.5 (635.4) &    0.4 &       553.2  &    13.1  &       312.2  &    20.6  &         0.5  &         0.9   \\
 1155+5453  &     11:55:31.7  &  $+$54:53:56.1  & 2010 Nov. 25 & 796.3 (8.3) &    0.3 &       199.7  &     7.5  &        60.6  &    20.0  &         0.5  &         1.8   \\
 1203+1319  &     12:03:20.7  &  $+$13:19:31.4  & 2011 Mar. 22 & 110.3  &    0.2  &   ($<$13.2)  &       -  &   ($<$31.3)  &       -  &    ($>$0.8)  &   ($>$-1.3)   \\
 1208+2514  &     12:08:05.6  &  $+$25:14:14.3  & 2011 Feb. 15 & 276.8 (60.0) &    0.2  &        94.4  &     3.5  &  ($<$133.5)  &       -  &         0.4  &   ($>$-0.5)   \\
 1210+3105  &     12:10:30.5  &  $+$31:05:18.7  & 2011 Mar. 27 &  31.6  &    0.1  &   ($<$20.8)  &       -  &   ($<$79.5)  &       -  &    ($>$0.2)  &   ($>$-2.0)   \\
 1210+3552  &     12:10:08.0  &  $+$35:52:39.4  & 2011 Mar. 27 &  20.9  &    0.1  &   ($<$10.9)  &       -  &   ($<$41.4)  &       -  &    ($>$0.2)  &   ($>$-2.0)   \\
 1213+5044  &     12:13:29.3  &  $+$50:44:29.4  & 2010 Dec. 30 & 102.7  &    0.1  &        76.7  &     2.3  &        95.5  &    26.4  &         0.1  &        -0.3   \\
 1220+2533  &     12:20:24.1  &  $+$25:33:38.1  & 2010 Dec. 28 &  26.6  &    0.1  &    ($<$7.5)  &       -  &   ($<$67.5)  &       -  &    ($>$0.5)  &   ($>$-3.3)   \\
 1222+0308  &     12:22:42.3  &  $+$03:08:30.5  & 2011 Feb. 19 &  12.9  &    0.1  &   ($<$14.4)  &       -  &  ($<$143.4)  &       -  &   ($>$-0.0)  &   ($>$-3.4)   \\
 1225+3214  &     12:25:13.1  &  $+$32:14:01.6  & 2010 Dec. 29 &  46.0  &    0.1  &        76.0  &     2.5  &        96.3  &    17.4  &        -0.2  &        -0.4   \\
 1230+4700  &     12:30:11.9  &  $+$47:00:22.7  & 2011 Jan. 02 &  87.5  &    0.1  &    ($<$7.1)  &       -  &   ($<$57.3)  &       -  &    ($>$0.9)  &   ($>$-3.1)   \\
 1243+0333  &     12:43:18.7  &  $+$03:33:00.6  & 2010 Dec. 30 &  64.5  &    0.1  &    ($<$8.6)  &       -  &   ($<$73.4)  &       -  &    ($>$0.7)  &   ($>$-3.2)   \\
 1246+1153  &     12:46:33.8  &  $+$11:53:47.9  & 2011 Mar. 23 &  53.0  &    0.1  &   ($<$14.1)  &       -  &   ($<$77.1)  &       -  &    ($>$0.5)  &   ($>$-2.5)   \\
 1250+0013  &     12:50:27.4  &  $+$00:13:45.6  & 2011 Mar. 23 &  62.0  &    0.2  &        79.8  &     4.6  &   ($<$47.2)  &       -  &        -0.1  &    ($>$0.8)   \\
 1308+4344  &     13:08:37.9  &  $+$43:44:15.1  & 2011 Jan. 02 &  52.3  &    0.1  &        42.9  &     2.7  &   ($<$58.8)  &       -  &         0.1  &   ($>$-0.5)   \\
 1320+3308  &     13:20:14.7  &  $+$33:08:36.3  & 2010 Nov. 27 &  82.1  &    0.1  &   ($<$30.2)  &       -  &   ($<$65.3)  &       -  &    ($>$0.4)  &   ($>$-1.2)   \\
 1324+3622  &     13:24:51.4  &  $+$36:22:42.7  & 2010 Nov. 25 & 822.1 (139.3) &    0.1 &       245.1  &    10.4 &       181.4  &    18.0  &         0.4  &         0.4   \\
 1326+3647  &     13:26:02.4  &  $+$36:47:59.3  & 2010 Nov. 26 & 892.1 (-) &    0.2 &       169.3 &    10.2 &        46.4\tablefootmark{b}  &    13.3  &         0.6  &         1.9   \\
 1330+3232  &     13:30:42.5  &  $+$32:32:49.1  & 2011 Feb. 17 &  16.9  &    0.1  &    ($<$8.3)  &       -  &  ($<$120.8)  &       -  &    ($>$0.3)  &   ($>$-4.0)   \\
 1334+1751  &     13:34:05.9  &  $+$17:51:24.7  & 2011 Feb. 14 &  28.8  &    0.1  &   ($<$10.5)  &       -  &   ($<$72.4)  &       -  &    ($>$0.4)  &   ($>$-2.9)   \\
 1334+3446  &     13:34:35.1  &  $+$34:46:40.0  & 2011 Feb. 18 &  17.4  &    0.1  &    ($<$8.6)  &       -  &   ($<$78.3)  &       -  &    ($>$0.3)  &   ($>$-3.3)   \\
 1336+0319  &     13:36:21.2  &  $+$03:19:51.2  & 2011 Jan. 01 &  33.2  &    0.1  &    ($<$8.9)  &       -  &   ($<$61.3)  &       -  &    ($>$0.5)  &   ($>$-2.9)   \\
 1337+1558  &     13:37:37.5  &  $+$15:58:20.0  & 2011 Mar. 26 &  30.0  &    0.1  &   ($<$12.1)  &       -  &   ($<$52.8)  &       -  &    ($>$0.3)  &   ($>$-2.2)   \\
 1350+3342  &     13:50:36.0  &  $+$33:42:17.4  & 2011 Jan. 02 &  99.7  &    0.1  &        88.5  &     2.5  &        44.9\tablefootmark{b}  &    10.8  &         0.0  &         1.0   \\
 1400+1751  &     14:00:26.4  &  $+$17:51:33.3  & 2011 Mar. 22 &  97.3  &    0.1  &        38.5  &     2.4  &   ($<$49.9)  &       -  &         0.3  &   ($>$-0.4)   \\
 1410+2338  &     14:10:43.4  &  $+$23:38:44.7  & 2011 Feb. 24 &  12.1  &    0.1  &    ($<$6.0)  &       -  &   ($<$67.1)  &       -  &    ($>$0.3)  &   ($>$-3.6)   \\
 1424+0239  &     14:24:47.4  &  $+$02:39:51.9  & 2011 Feb. 22 &  10.8  &    0.2  &    ($<$4.2)  &       -  &  ($<$142.1)  &       -  &    ($>$0.3)  &   ($>$-5.3)   \\
 1424+0244  &     14:24:36.3  &  $+$02:44:42.5  & 2011 Feb. 17 &  12.3  &    0.2  &    ($<$9.0)  &       -  &  ($<$119.1)  &       -  &    ($>$0.1)  &   ($>$-3.9)   \\
 1424+2637  &     14:24:40.5  &  $+$26:37:30.5  & 2010 Nov. 27 & 343.4 (12.0) &    0.2 &        81.5  &     8.6  &   ($<$32.9)  &       -  &         0.5  &    ($>$1.4)   \\
 1429+0715  &     14:29:55.4  &  $+$07:15:12.9  & 2010 Dec. 30 &  45.8  &    0.2  &        58.4\tablefootmark{b}  &     4.4  &   ($<$86.0)  &       -  &        -0.1  &   ($>$-0.6)   \\
 1436+0519  &     14:36:20.4  &  $+$05:19:51.5  & 2010 Dec. 31 &  26.7  &    0.1  &    ($<$9.9)  &       -  &   ($<$52.5)  &       -  &    ($>$0.4)  &   ($>$-2.5)   \\
 1447+1340  &     14:47:02.1  &  $+$13:40:06.2  & 2011 Feb. 25 &  10.8  &    0.1  &   ($<$11.4)  &       -  &  ($<$136.6)  &       -  &   ($>$-0.0)  &   ($>$-3.7)   \\
 1450+1006  &     14:50:49.4  &  $+$10:06:49.1  & 2011 Jan. 02 &  50.0 (47.9) &    0.1  &        52.7  &     1.0  &       102.3  &    15.0  &        -0.0  &        -1.0   \\
 1452+1654  &     14:52:43.3  &  $+$16:54:13.5  & 2011 Feb. 18 &  19.2  &    0.1  &    ($<$7.0)  &       -  &   ($<$63.1)  &       -  &    ($>$0.4)  &   ($>$-3.3)   \\
 1454+1838  &     14:54:31.5  &  $+$18:38:32.3  & 2010 Dec. 29 &  37.7  &    0.2  &   ($<$11.1)  &       -  &   ($<$70.9)  &       -  &    ($>$0.4)  &   ($>$-2.8)   \\
 1501+1742  &     15:01:52.3  &  $+$17:42:28.2  & 2011 Feb. 20 &  17.0  &    0.1  &   ($<$11.9)  &       -  &  ($<$168.5)  &       -  &    ($>$0.1)  &   ($>$-4.0)   \\
 1504+2600  &     15:04:57.1  &  $+$26:00:58.5  & 2010 Nov. 26 & 228.9 (63.3) &    0.3 &        47.6  &     7.5  &   ($<$41.7)  &       -  &         0.6  &    ($>$0.2)   \\
 1506+1250  &     15:06:56.4  &  $+$12:50:48.6  & 2011 Mar. 22 &  75.9 (70.0) &    0.1 &       100.5  &     2.5  &   ($<$57.3)  &       -  &        -0.1  &    ($>$0.8)   \\
 1508$-$0011&     15:08:53.9  &  $-$00:11:49.0  & 2011 Mar. 27 &  23.0  &    0.2  &   ($<$11.5)  &       -  &   ($<$85.1)  &       -  &    ($>$0.3)  &   ($>$-3.0)   \\
 1516+0015  &     15:16:40.2  &  $+$00:15:01.8  & 2011 Mar. 29 & 1090.2 (754.1) &    0.3  &      1121.9  &     8.0  &      1003.6  &    37.9  &        -0.0  &         0.2   \\
 1530+2705  &     15:30:16.2  &  $+$27:05:51.0  & 2011 Feb. 23 &  13.4  &    0.1  &   ($<$8.8)  &     -  &   ($<$80.7)   &    - &     ($>$0.2)  &        ($>$-3.3)   \\
 1535+1347  &     15:35:35.1  &  $+$13:47:52.8  & 2011 Mar. 25 &  33.5  &    0.2  &   ($<$10.3)  &       -  &   ($<$54.1)  &       -  &    ($>$0.4)  &   ($>$-2.5)   \\
 1539+5530  &     15:39:35.6  &  $+$55:30:15.9  & 2011 Jan. 01 &  31.5  &    0.1  &    ($<$9.9)  &       -  &   ($<$71.8)  &       -  &    ($>$0.4)  &   ($>$-3.0)   \\
 1544+4700  &     15:44:26.9  &  $+$47:00:24.2  & 2011 Feb. 21 &  17.4  &    0.1  &   ($<$13.6)  &       -  &   ($<$94.9)  &       -  &    ($>$0.1)  &   ($>$-2.9)   \\
 1559+4442  &     15:59:54.0  &  $+$44:42:32.4  & 2010 Dec. 31 &  58.8  &    0.2  &   ($<$10.4)  &       -  &   ($<$54.3)  &       -  &    ($>$0.6)  &   ($>$-2.5)   \\
 1604+1744  &     16:04:26.5  &  $+$17:44:31.2  & 2011 Mar. 23 &  71.8 (71.4) &    0.1 &       117.2  &     4.0  &       105.0  &    11.9  &        -0.2  &         0.2   \\
 1608+2828  &     16:08:21.1  &  $+$28:28:43.3  & 2010 Nov. 28 & 108.2 (77.3) &    0.1 &        97.8  &     9.4  &        84.7  &    22.5  &         0.0  &         0.2   \\
 1617+3500  &     16:17:40.5  &  $+$35:00:15.1  & 2010 Nov. 27 & 365.4 (59.7) &    0.2 &       110.0  &     8.2  &   ($<$70.1)  &       -  &         0.4  &    ($>$0.7)   \\
 1620+2400  &     16:20:47.1  &  $+$24:00:51.1  & 2010 Dec. 31 &  28.5  &    0.2  &   ($<$11.2)  &       -  &   ($<$79.6)  &       -  &    ($>$0.3)  &   ($>$-2.9)   \\
 1624+4831  &     16:24:24.5  &  $+$48:31:42.4  & 2010 Dec. 29 &  69.0  &    0.1  &   ($<$12.6)  &       -  &   ($<$40.6)  &       -  &    ($>$0.6)  &   ($>$-1.8)   \\
 1628+2529  &     16:28:46.1  &  $+$25:29:40.9  & 2010 Dec. 27 &  27.5  &    0.1  &    ($<$9.9)  &       -  &   ($<$92.2)  &       -  &    ($>$0.4)  &   ($>$-3.3)   \\
 1658+2523  &     16:58:30.1  &  $+$25:23:24.9  & 2011 Feb. 20 &  14.8  &    0.1  &   ($<$10.5)  &       -  &   ($<$96.8)  &       -  &    ($>$0.1)  &   ($>$-3.3)   \\
 1703+2410  &     17:03:58.5  &  $+$24:10:39.6  & 2010 Dec. 31 &  22.8  &    0.1  &   ($<$12.7)  &       -  &   ($<$72.8)  &       -  &    ($>$0.2)  &   ($>$-2.6)   \\
 1715+5724  &     17:15:23.0  &  $+$57:24:40.3  & 2010 Dec. 31 &  49.3  &    0.1  &        38.7\tablefootmark{b}  &     1.1  &        68.5\tablefootmark{b}  &    14.6  &         0.1  &        -0.9   \\
\end{longtable}
\tablefoot{
Flux at 1.4 GHz in parentheses indicates core flux of galaxies with extended features.
\tablefoottext{a}{Flux and error at 1.4 GHz are estimated from the FIRST survey. In the case of multiple radio detections, we added all the fluxes corresponding to the galaxy.}
\tablefoottext{b}{When one of the lines (either AZ or EL) showed radio detection, the other line did not. Thus, pointing correction has not been done.}
}
\end{longtab}

\end{document}